\documentclass[submitting]{nst}

\usepackage{subfigure,dcolumn}
\usepackage[T2A,T1]{fontenc}
\usepackage[russian,english]{babel}

\usepackage{listings}
\def\NvDEx{N$\nu$DEx}
\def\82SeF6{$^{82}$SeF$_6$}
\def\Se82{$^{82}$Se}
\def\SeF6{SeF$_6$}
\def\SF6{SF$_6$}
\def\0vbb{0$\nu\beta\beta$}
\def\2vbb{2$\nu\beta\beta$}

\begin{document}

\title{\boldmath N$\nu$DEx-100  Conceptual Design Report}\thanks{Supported by the National Key Research and Development Program of China (No. 2021YFA1601300 and 2022YFA1604703), From-0-to-1 Original Innovation Program of Chinese Academy of Sciences (No. ZDBS-LY-SLH014), International Partner Program of Chinese Academy of Sciences (No. GJHZ2067), and National Natural Science Foundation of China Youth Science Fund Project (No. 12105110)}

% authors
\author{X. Cao}
\affiliation{Shanghai Advanced Research Institute, Chinese Academy of Sciences, Shanghai 201210, China}

\author{Y. Chang}
\affiliation{Lanzhou University, Lanzhou 730000, China}

\author{K. Chen}
\affiliation{Central China Normal University, Wuhan 430079, China}

\author{E. Ciuffoli}
\affiliation{Institute of Modern Physics, Chinese Academy of Sciences, Lanzhou 730000, China}

\author{L. Duan}
\affiliation{Institute of Modern Physics, Chinese Academy of Sciences, Lanzhou 730000, China}

\author{D. Fang}
\affiliation{Institute of Modern Physics, Chinese Academy of Sciences, Lanzhou 730000, China}

\author{C. Gao}

\author{S. K. Ghorui}
\affiliation{Institute of Modern Physics, Chinese Academy of Sciences, Lanzhou 730000, China}

\author{P. Hu}
\affiliation{Lanzhou University, Lanzhou 730000, China}

\author{Q. Hu}
\affiliation{Institute of Modern Physics, Chinese Academy of Sciences, Lanzhou 730000, China}

\author{S. Huang}
\affiliation{Institute of Modern Physics, Chinese Academy of Sciences, Lanzhou 730000, China}

\author{Z. Huang}
\affiliation{Lanzhou University, Lanzhou 730000, China}

\author{L. Lang}
\affiliation{Central China Normal University, Wuhan 430079, China}

\author{Y. Li}
\affiliation{Shanghai Institute of Applied Physics, Chinese Academy of Sciences, Shanghai 201800, China}

\author{Z. Li}
\affiliation{Institute of Modern Physics, Chinese Academy of Sciences, Lanzhou 730000, China}

\author{T. Liang}
\affiliation{Central China Normal University, Wuhan 430079, China}

\author{J. Liu}
\affiliation{Central China Normal University, Wuhan 430079, China}

\author{C. Lu}
\affiliation{Institute of Modern Physics, Chinese Academy of Sciences, Lanzhou 730000, China}

\author{F. Mai}
\affiliation{Institute of Modern Physics, Chinese Academy of Sciences, Lanzhou 730000, China}

\author{Y. Mei}
\affiliation{Lawrence Berkeley National Laboratory, Berkeley, CA 94720, USA}

\author{H. Qiu}
\email[Corresponding author, ]{Hao Qiu, Institute of Modern Physics, Chinese Academy of Sciences, Lanzhou 730000, China, 17797625768, qiuh@impcas.ac.cn.}
\affiliation{Institute of Modern Physics, Chinese Academy of Sciences, Lanzhou 730000, China}

\author{X. Sun}
\affiliation{Central China Normal University, Wuhan 430079, China}

\author{X. Tang}
\affiliation{Shanghai Institute of Applied Physics, Chinese Academy of Sciences, Shanghai 201800, China}

\author{H. Wang}
\affiliation{Central China Normal University, Wuhan 430079, China}

\author{Q. Wang}
\affiliation{Lanzhou University, Lanzhou 730000, China}

\author{L. Xiao}
\affiliation{Central China Normal University, Wuhan 430079, China}

\author{M. Xiao}
\affiliation{Institute of Modern Physics, Chinese Academy of Sciences, Lanzhou 730000, China}

\author{J. Xin}
\affiliation{Lanzhou University, Lanzhou 730000, China}

\author{N. Xu}
\affiliation{Institute of Modern Physics, Chinese Academy of Sciences, Lanzhou 730000, China}
\affiliation{Lawrence Berkeley National Laboratory, Berkeley, CA 94720, USA}

\author{P. Yang}
\affiliation{Institute of Modern Physics, Chinese Academy of Sciences, Lanzhou 730000, China}

\author{Y. Yang}
\affiliation{Institute of Modern Physics, Chinese Academy of Sciences, Lanzhou 730000, China}

\author{Z. Yang}
\affiliation{Institute of Modern Physics, Chinese Academy of Sciences, Lanzhou 730000, China}

\author{Z. Yu}
\affiliation{Central China Normal University, Wuhan 430079, China}

\author{D. Zhang}
\affiliation{Central China Normal University, Wuhan 430079, China}

\author{J. Zhang}
\affiliation{North China University of Water Resources and Electric Power, Zhengzhou 450045, China}

\author{C. Zhao}
\affiliation{Institute of Modern Physics, Chinese Academy of Sciences, Lanzhou 730000, China}

\author{D. Zhu}
\affiliation{Central China Normal University, Wuhan 430079, China}

\begin{abstract}
Observing nuclear neutrinoless double beta (\0vbb) decay would be a revolutionary result in particle physics. Observing  such a decay would prove that the neutrinos are their own antiparticles, help to study the absolute mass of neutrinos, explore the origin of their mass, and may explain the matter-antimatter asymmetry in our universe by lepton number violation.

We propose developing a time projection chamber (TPC) using high-pressure \82SeF6 \ gas and top-metal silicon sensors for read-out in the China Jinping Underground Laboratory (CJPL) to search for neutrinoless double beta decay of \Se82, called the \NvDEx \ experiment.
Besides being located at CJPL with the  world's thickest rock shielding, \NvDEx \ combines the advantages of the high $Q_{\beta\beta}$ (2.996 MeV) of \Se82 \ and the TPC's ability to distinguish signal and background events using their different topological characteristics.
This makes \NvDEx \ unique, with great potential for low-background and high-sensitivity \0vbb~ searches.

\NvDEx-100, a \NvDEx \ experiment phase with 100 kg of \SeF6 gas, is being built, with plans to complete installation at CJPL by 2025.
This report introduces \0vbb \ physics, the \NvDEx \ concept and its advantages, and the schematic design of \NvDEx-100, its subsystems, and background and sensitivity estimation.
\end{abstract}


\collaboration{\NvDEx-100 collaboration}

\maketitle

\section{The Physics}
\label{sec:physics}

The Standard Model (SM) of particle physics is an important cornerstone of physics and the entire natural sciences that has successfully undergone experimental testing for more than half a century. 
The discovery of its last component, the Higgs particle, marked the perfect end of an era. In the SM, neutrinos have no mass. However, their oscillations, which have been observed nowadays by many independent experiments and are supported by irrefutable evidence, require the presence of a non-diagonal mass term in the flavor basis. This is the first experimental proof of physics beyond the SM in particle physics.
Some properties of neutrinos remain unknown, such as whether they are Dirac or Majorana fermions, their absolute mass, and their mass hierarchy.

The charged fermions in the SM are all Dirac particles, which gain mass through Yukawa coupling with the Higgs boson. 
Because neutrinos are electrically neutral, they are the only candidates in the SM to be Majorana fermions, {\it i.e.} they could be their own antiparticles. 
If this is the case, we can also explain why their masses are significantly lower than those of other charged leptons in the SM by introducing a seesaw mechanism~\cite{Mohapatra:1979ia}.
Neutrinoless double beta decay experiments are the ideal way to determine if this is the case; if such a process is observed, it would be irrefutable proof that neutrinos are Majorana particles, opening the door to new physics. 
The measured decay rate can quantitatively constrain the absolute mass and mass ordering of neutrinos. Additionally, neutrinoless double beta decay violates lepton number and CP parity conservations, which can generate a net lepton number in the early universe evolution, thus explaining the matter-antimatter asymmetry in the universe.

\section{\NvDEx \ Concept and its Advantages}
\label{sec:concept}

The rate of neutrinoless double beta (\0vbb) decay occurrence (if it occurs) is extremely low, making experimental observations difficult. These  experiments have been developed for decades, with intense competition among the various experimental approaches. Existing large-scale experiments include GERDA~\cite{GERDA:2020xhi}, MAJORANA~\cite{Majorana:2017csj}, CUORE~\cite{CUORE:2021mvw}, CUPID~\cite{Augier:2022znx}, KamLAND-Zen~\cite{KamLAND-Zen:2022tow}, and EXO~\cite{EXO-200:2019rkq}.
In China, experiments including CDEX~\cite{CDEX:2017pgl}, PandaX~\cite{PandaX-II:2019euf}, CUPID-China~\cite{Xue:2017qbf}, and JUNO~\cite{Zhao:2016brs} have searched for or are being developed to search for \0vbb decay.
Currently, the highest experimental half-life sensitivity reaches $10^{25}-10^{26}$ years, yet no such decay has been observed. Next-generation \0vbb decay experiments are approaching the sensitivity needed for the inverted hierarchy of neutrino masses, on the order of $10^{27}$ years for most \0vbb decay isotopes. For the normal hierarchy of neutrino masses, which is slightly favored by oscillation experiment results, the required experimental half-life sensitivity is two orders of magnitude higher, on the order of about $10^{29}$ years.

Reducing the experimental background is key to improving the sensitivity of neutrinoless double beta decay experiments. With zero background, the experiment’s sensitivity is proportional to the exposure (mass of decay isotope $\times$ experiment time). However, with a high background, the experimental sensitivity increases with the square root of exposure\cite{Gomez-Cadenas:2010zcc}. Thus, to increase experimental sensitivity by another 1-3 orders of magnitude, innovative techniques must be applied to significantly reduce the experimental background.

The concept of a “No neutrino Double-beta-decay Experiment (\NvDEx)", searching for the neutrinoless double beta decay of \Se82 using a high-pressure gas time projection chamber (TPC) with \82SeF6 as the working medium and read out by Topmetal sensor chips, was proposed by D.R. Nygren, B.J.P. Jones, N. López-March, Y. Mei, F. Psihas and J. Renner in 2018~\cite{Nygren:2018ewr}. This scheme combines the high $Q_{\beta\beta}$ of \Se82 with the ability of TPC to distinguish between the signal and background using event topology, which can significantly reduce the experimental background. The $Q_{\beta\beta}$ of \Se82 is as high as 2.996 MeV, which is higher than most natural radioactive backgrounds and that of the decay isotopes currently used in many mainstream experiments. For example, the natural radioactive $\gamma$ background near the $Q_{\beta\beta}$ of \Se82 is more than two orders of magnitude lower than that around the $Q_{\beta\beta}$ of $^{136}Xe$ (2.458 MeV). Meanwhile, in gaseous TPC, the double beta decay can be reconstructed as two electron tracks, each with a distinct Bragg peak at the end. This feature can be used to distinguish the signal from the background.

However, this experimental concept faces a major technical challenge: \SeF6 is an electronegative gas, in which the electrons generated by ionization quickly combine with gas molecules to form negative ions, and electron avalanche amplification cannot happen. Thus, the weak signals cannot be read out with traditional technologies. To solve this problem, we designed the Topmetal-S sensor\cite{Gao:2019ohr,You:2021yqk}, a kind of silicon sensor chip with a layer of metal on top, dedicated to \0vbb decay experiments, making TPC without physical amplification possible. It adopts an industrial semiconductor CMOS process, and the top layer has a metal sheet for charge collection. In principle, its noise level can be as low as about 30 e$^-$; thus, the primary ionized charge can be directly read out without physical amplification, providing a unique opportunity to search for \0vbb decay using the \82SeF6 gas TPC.

The construction of the China Jinping Underground Laboratory (CJPL) provides a unique opportunity for developing  \0vbb decay experiments. CJPL has the thickest natural rock shield in the world, and the second phase of CJPL is being constructed with a world-class experimental space and a low background environment. \NvDEx \ will be developed at CJPL, taking full advantage of its low background level and large space.

\section{\NvDEx-100 Schematic Design}
\label{sec:design}

\subsection{\NvDEx-100 Overall Design}
\label{sec:overallDesign}

\begin{figure}[ht]
    \centering
    \includegraphics[width=\linewidth]{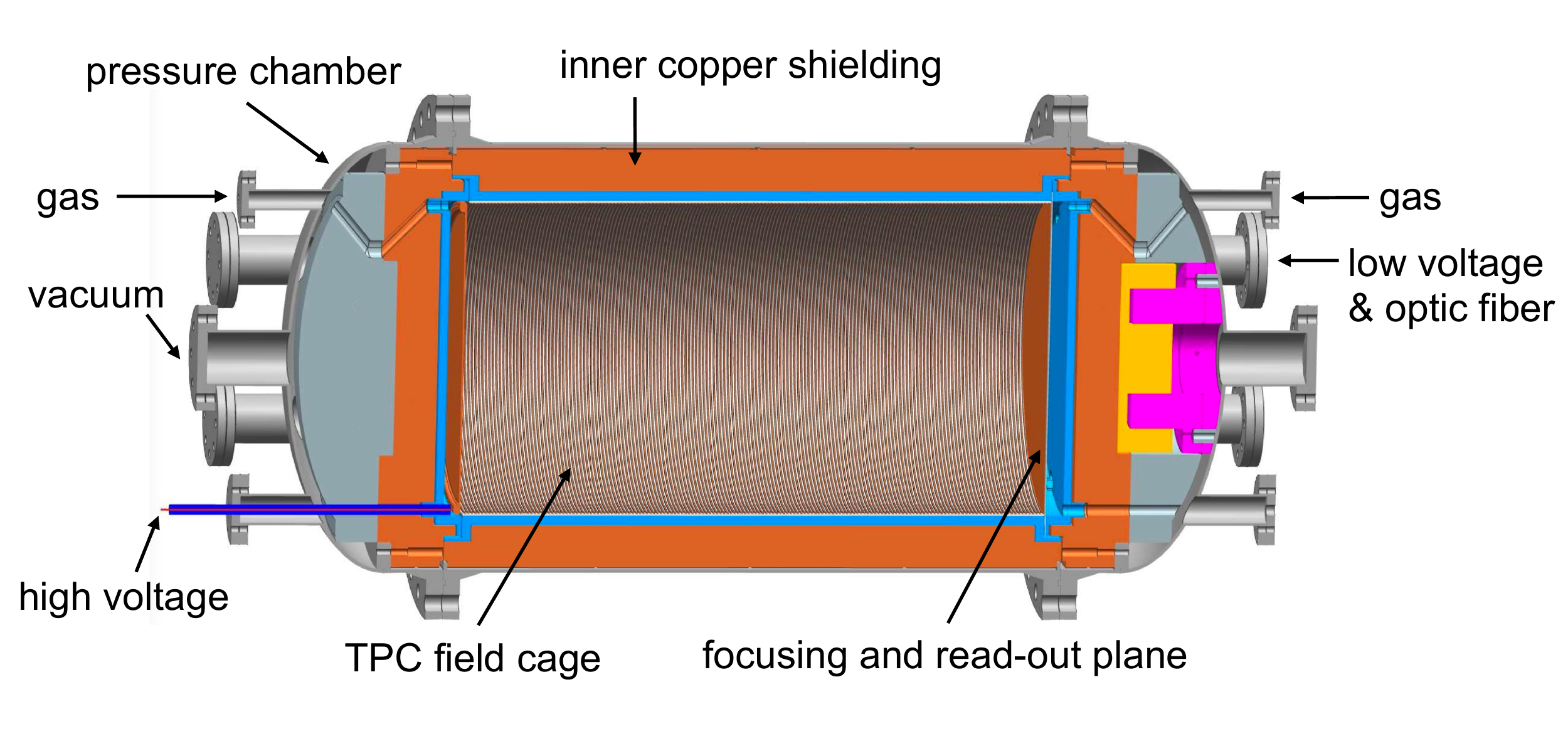}
    \caption{Schematic design of the main part of the \NvDEx-100 experiment. 
}
    \label{fig:NvDEx-100}
\end{figure}

This study develops a \NvDEx-100 experiment with 100kg of natural \SeF6 gas. The preliminary design is shown in Fig. \ref{fig:NvDEx-100}. The main body of the experiment is in a pressure chamber, with feed-through flanges for gas, low-voltage, optical fibers, and high-voltage. An inner copper shielding inside the pressure chamber shields most of the external radiation. The core detector of the experiment - TPC - is installed in the barrel part of the chamber, comprising an insulating layer, a high-voltage plane, a field cage, and a readout plane. The readout plane consists of the focusing layer and the readout electronics layer on which the Topmetal-S sensor chips are mounted. In addition to the main body of the experiment,  there are lead and high-density polyethylene (HDPE) external shieldings surrounding the pressure chamber, as well as auxiliary facilities such as the gas system, which are not shown in Fig. \ref{fig:NvDEx-100}.

During the experiment, \Se82 (with a natural abundance of 8.7\%) in \SeF6 gas undergoes double beta decay, releasing two electrons. The total energy of the two electrons is 2.996 MeV for the neutrinoless double beta decay. These two electrons lose energy in the gas, ionize the gas, and form curved tracks due to scattering. At the ends of the two tracks, two Bragg peaks with the largest energy loss are formed. Due to the electronegativity of \SeF6, the electrons generated by ionization quickly form negative ions with surrounding \SeF6 molecules. Finally, various SeF$_{N}^{\pm}$ ions are formed in certain fractions, including SeF$_{0-5}^{+}$ and SeF$_{5,6}^{-}$, which drift to the two ends of the TPC in the electric field. After the SeF$_{5,6}^{-}$ ions reach the readout plane, their signals are read out. The drift velocities of SeF$_{5}^{-}$ and SeF$_{6}^{-}$ ions are different, as are their arrival times. This time difference can be used to obtain the drift distance. The readout plane comprises the focusing and readout electronics layers; the focusing layer generates a certain electric field structure with small holes that allow the passage and collection of drift charges with ~100\% efficiency at the 1 mm$^2$-sized readout electrodes on the surface of the Topmetal-S chips. The Topmetal-S chips, located on the surface of the readout electronics layer, measure the charge and time of the signal and generate digital data, which are collected by the electronic readout boards and transmitted to the data acquisition computer through the optical fibers.

\subsection{Pressure Chamber and Inner Copper Shielding}
\label{sec:pressureChamber}

\begin{figure}
    \centering
    \includegraphics[width=\linewidth]{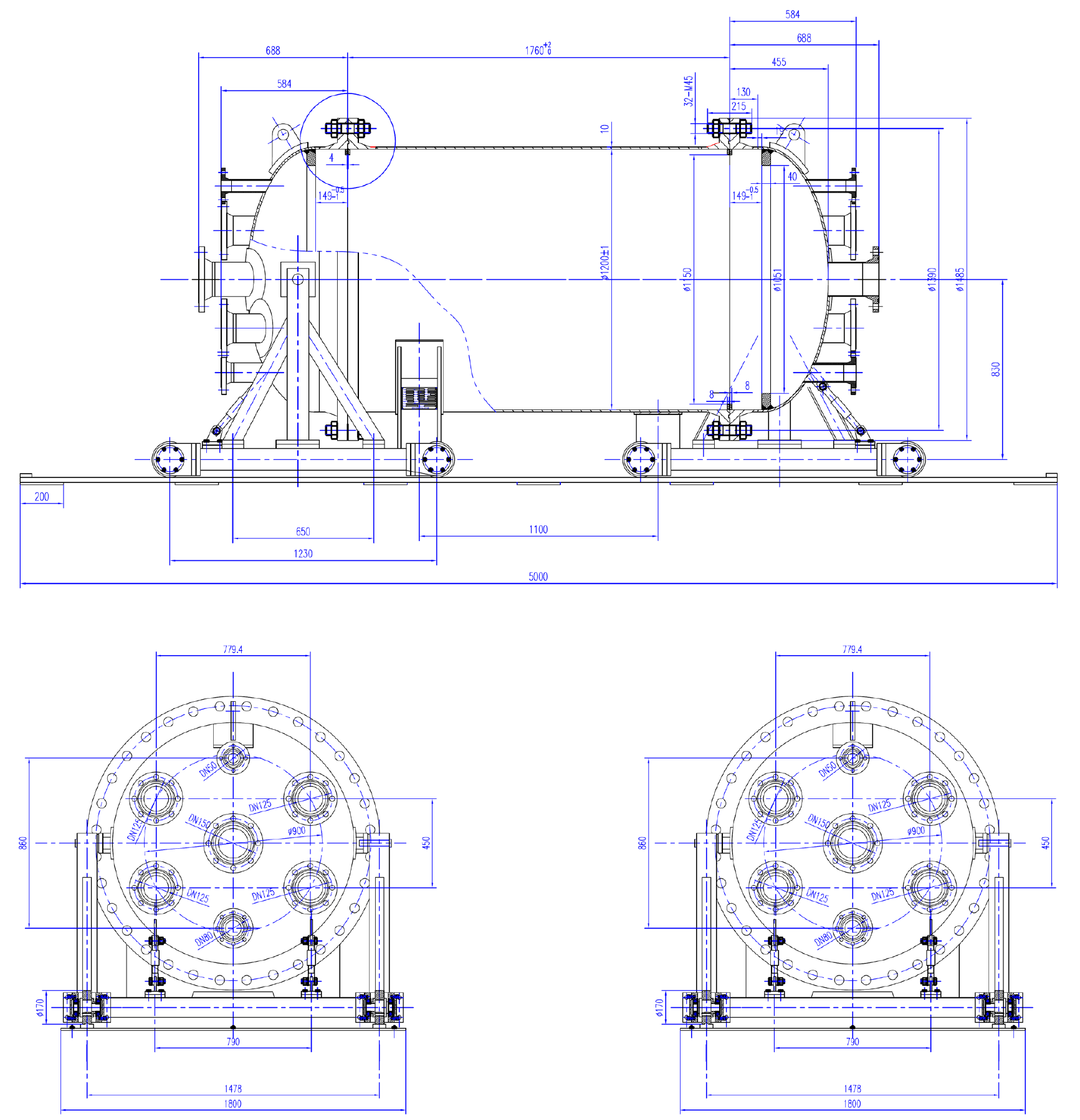}
    \caption{Design of the pressure chamber.}
    \label{fig:NvDEx-HPV-CAD}
\end{figure}

\begin{figure}[ht]
    \centering
    \includegraphics[width=\linewidth]{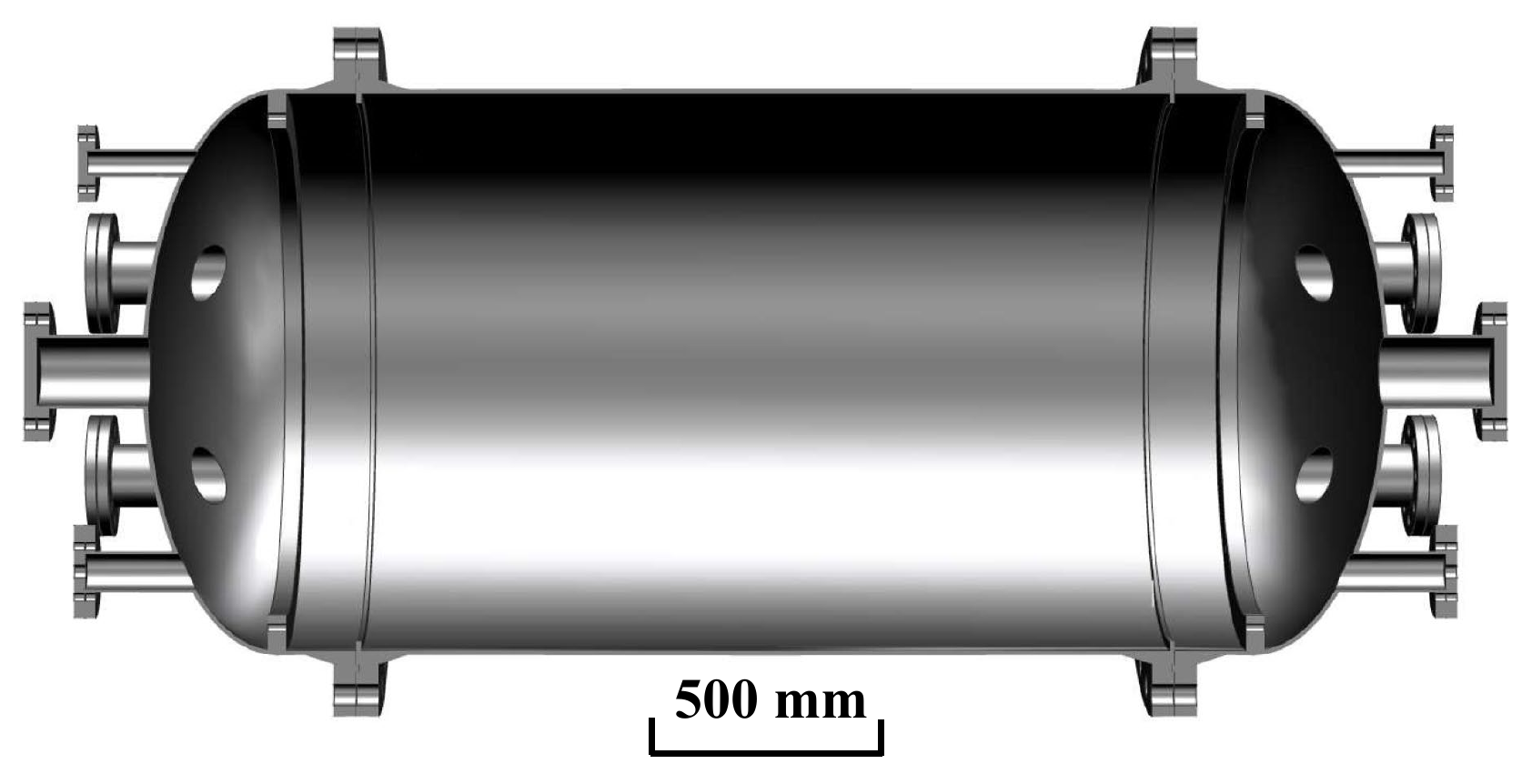}
    \caption{Cross-sectional view of the pressure chamber.}
    \label{fig:NvDEx-HPV-chamber}
\end{figure}

\NvDEx-100 uses the \SeF6 \ gas at a pressure of 1.0 MPa. This pressure is chosen to obtain more \SeF6~ gas mass within a certain volume while avoiding liquefication~\cite{Nygren:2018ewr}. The pressure chamber design is shown in Fig. \ref{fig:NvDEx-HPV-CAD}. The chamber comprises a barrel and two end caps, connected  with two DN1200 Tongue-Grove (T-G) flanges. There are seven smaller T-G flanges  on each end cap: one DN50 flange for gas, one DN80 flange for high voltage, four DN125 flanges for low-voltage and optic fibers, and one DN150 flange for vacuum. The inner diameter and length of the barrel are 1200 mm and 1760 mm, respectively. The chamber is made of 10 mm-thick, low background, stainless steel. Figure \ref{fig:NvDEx-HPV-chamber} shows the cross-sectional view of the chamber. The weight of the chamber is around 2211 kg without considering the bolts. The barrel part of the pressure chamber sits on two saddles, while the two end caps are supported with carts, which can move away along the rails when opening the chamber. 

\begin{figure}[ht]
    \centering
    \includegraphics[width=0.8\linewidth]{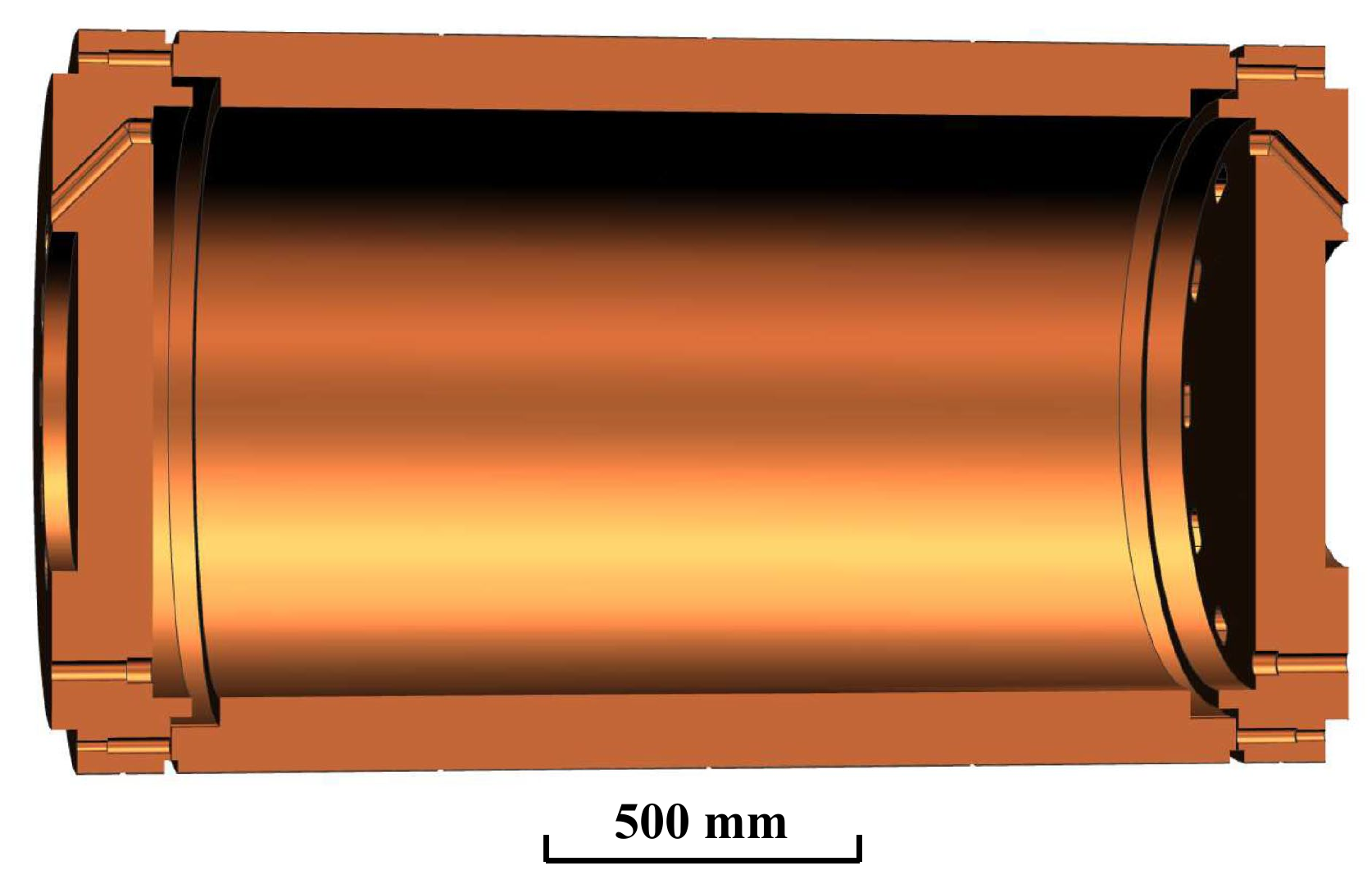}
    \caption{Cross-sectional view of the designed inner copper shielding.}
    \label{fig:NvDEx-copper-shield-cs}
\end{figure}

  A 12 cm-thick oxygen-free  copper shielding with low radioactive isotope contamination is placed inside the pressure chamber to suppress the background radiation from the pressure chamber and outside, as shown in Fig. \ref{fig:NvDEx-100}. Figure \ref{fig:NvDEx-copper-shield-cs} shows a cross-sectional view of the inner copper shielding design, comprising a barrel part and two disks mounted in the end cups of the pressure chamber. The outer and inner diameters of the barrel part are 1190 mm and 950 mm, respectively. The barrel part and the disks weigh about 6108 kg and 1476 kg, respectively. The disks have some holes enabling the passage of gas, optic fibers, low-voltage cables, and the high-voltage feedthrough. The holes, except the ones for high voltage, are tilted to avoid outside radiation reaching the sensitive volume of the TPC via a straight path.

\begin{figure}[ht]
    \centering
    \includegraphics[width=0.4\linewidth]{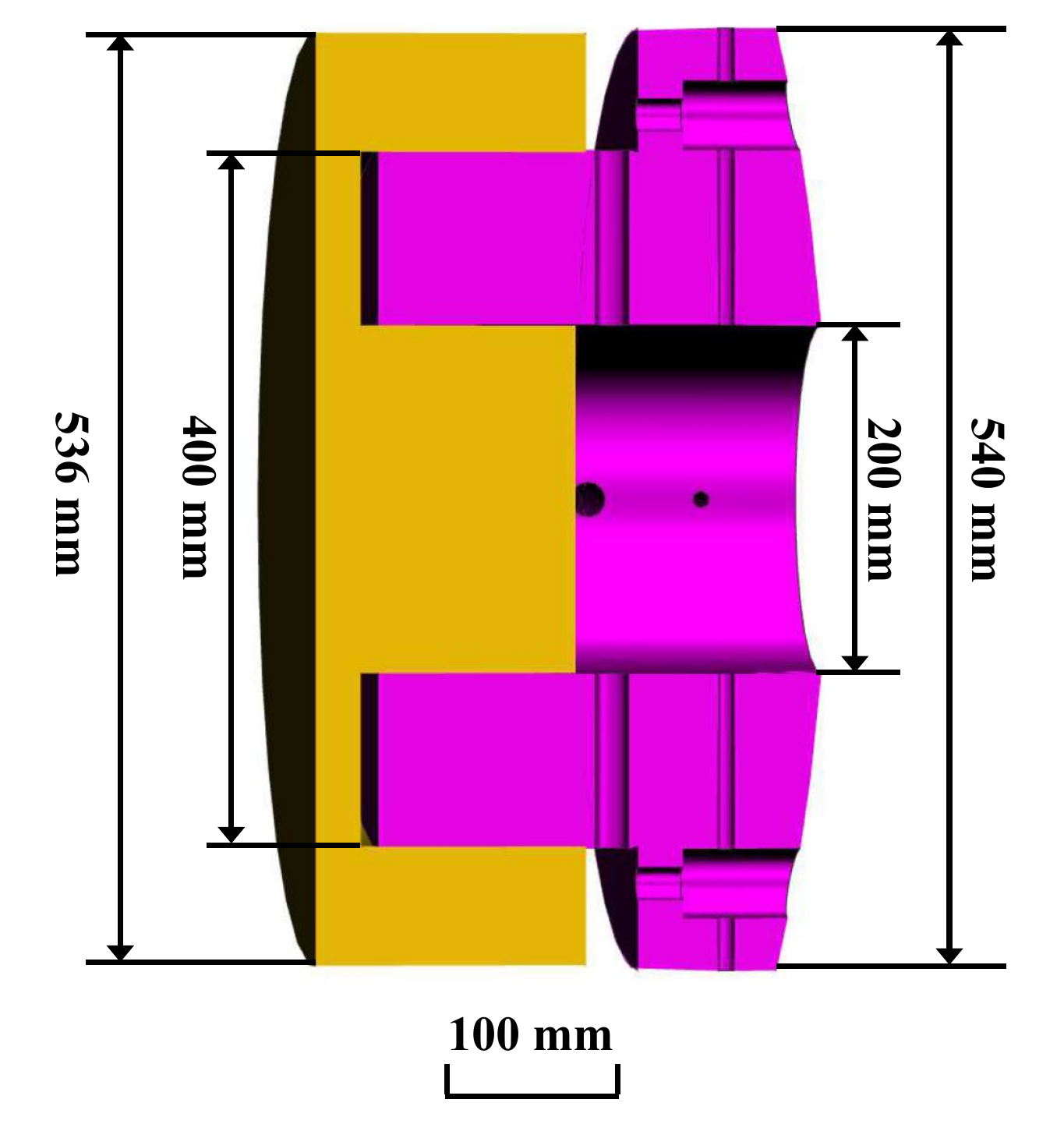}
    \caption{Cross-sectional view of the cooling base and tube.}
    \label{fig:NvDEx-cooling-tube-cs}
\end{figure}

The Topmetal sensors and electronics on the readout plane will generate heat when taking data, which will induce convection in the \SeF6 gas in the sensitive volume, with a maximum velocity larger than 10 cm/s assuming heat power of 700 W on the readout plane. This could be a problem for \NvDEx \ TPC because the ions drift very slowly, with a velocity above 20 cm/s, which is significantly lower than the drift velocity of electrons (on the order of several cm/$\mu$s in most other TPC's), and the convection could cause serious distortion of the reconstructed event topology.
Thus, cooling the readout plane and minimizing the temperature non-uniformity inside the TPC is necessary. For this purpose, a copper heat conductor will be placed between the inner copper shielding disk and the end cap of the pressure vessel. As shown in Fig. \ref{fig:NvDEx-100}, the heat conductor is composed of a base (in yellow) and a tube (in purple) fixed to the inner copper shielding disk and to the pressure chamber end cap, respectively. Figure \ref{fig:NvDEx-cooling-tube-cs} shows the cross-sectional view of the copper heat conductor with dimensions. The base and the tube can slide horizontally relative to each other. This design ensures good contact between all neighboring parts along the heat conduction path, even when the pressure chamber expands due to the gas pressure, so that the total heat resistance is acceptable.
The weight of the copper heat conductor is around 489 kg.
A liquid cooling plate will be mounted on the outer surface of the end cup of the pressure chamber. The temperature difference in the TPC and convection in the gas will be minimized by adjusting the temperature of the cooling plate.

\begin{figure}[ht]
    \centering
    \includegraphics[width=0.5\linewidth]{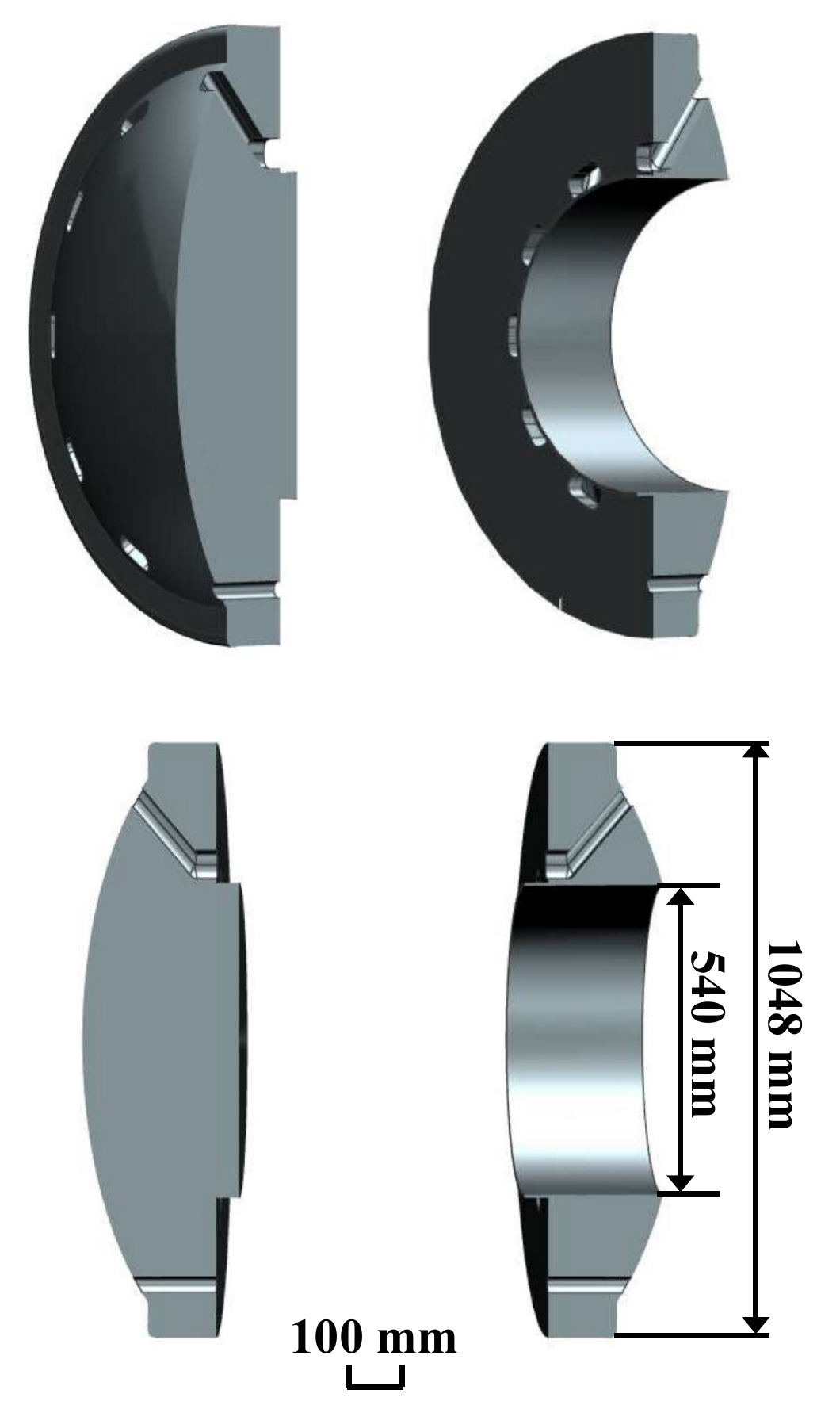}
    \caption{Cross-sectional view of the POM fillers.}
    \label{fig:NvDEx-POM-filler-cs}
\end{figure}

The \SeF6 and \82SeF6 gases are very expensive. Two plastic fillers will be placed in the end caps of the pressure chamber to reduce the amount of the gas to be used, occupying the gap outside the inner copper shielding disks, as shown in Fig. \ref{fig:NvDEx-100}.
Since \SeF6 is toxic, any material that absorbs the gas and gradually releases it when the pressure chamber is open during the maintenance of the experiment could endanger people and the environment.
Considering this, the fillers, as well as the insulator layer and the TPC field cage supporting cylinder to be described in the next subsection, will be made of polyoxymethylene (POM), which absorbs the minimum amount of gas among plastic materials with acceptable mechanical strength.
The design of the fillers is shown in Fig. \ref{fig:NvDEx-POM-filler-cs}. There are also some holes in the fillers for gas, optic fibers, low voltage cables, and the high voltage feedthrough to go through.

The pressure chamber and the inner copper shielding have been manufactured for an above-ground prototype experiment. The copper heat conductor and the fillers are being manufactured. The above-ground prototype will be assembled in the near future.
Then, tests of the gas tightness of the pressure chamber, heat conduction, and temperature control of the readout plane will be conducted. 

\subsection{Field Cage}
\label{sec:fieldCage}

The electronegativity of the \SeF6 gas used in \NvDEx-100 is very high. This means the negatively charged particles drifting towards the readout plane will not be electrons since they are quickly captured, but negative ions instead.
The readout plane will employ innovative Topmetal-S sensors to read out the drifted charge without physical amplification like an electron avalanche. Details about Topmetal-S sensors and the readout plane will be introduced in Sub-section \ref{sec:Topmetal}.

Most of the drift negative ions will be SeF$_{6}^{-}$ and SeF$_{5}^{-}$; however many more complex molecules may be formed. The drifting negative ions may form clusters like SeF$_{6}^{-}$(SeF$_6$)$_{n}$ and SeF$_{5}^{-}$(SeF$_6$)$_{n}$
(n=1,2,3,...) with a low drift field. These clusters will smear the drift velocity of the negative ions, resulting in increased noise. Similar to \SF6, the cluster formation in \SeF6 can be suppressed with high drift fields.
Consequently, the drift field of \NvDEx-100 will be as high as 400 V/cm, corresponding to a drift velocity of negative ions above 20 cm/s. 

\begin{figure}
    \centering
    \includegraphics[width=\linewidth]{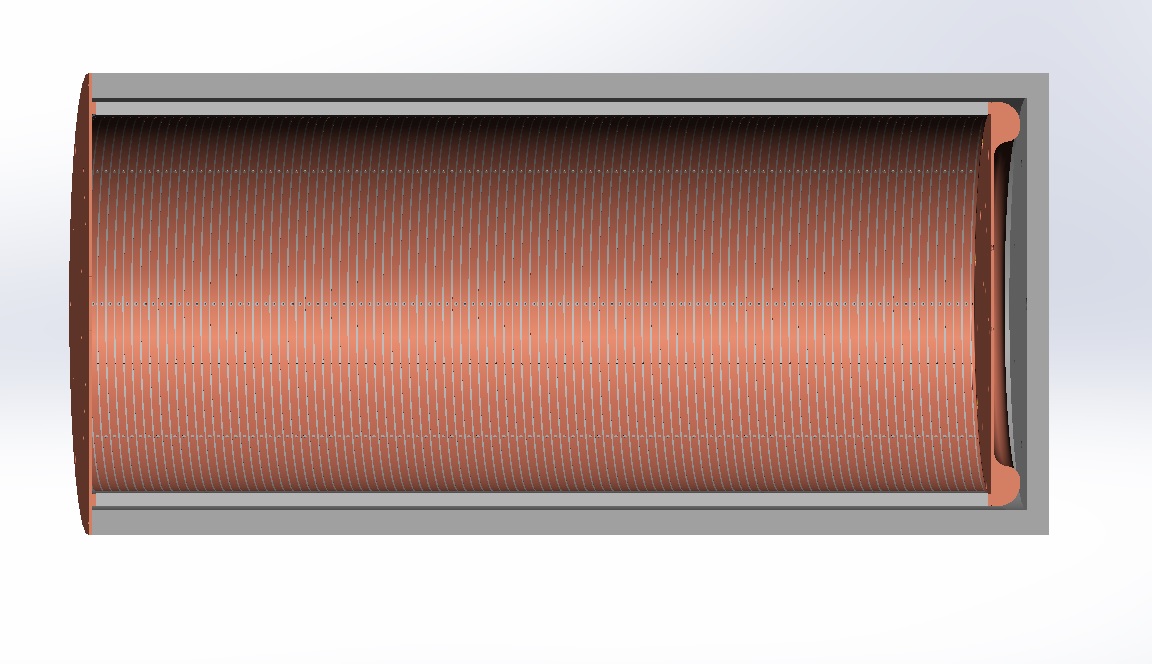}
    \caption{Cross-sectional view of the design of a prototype field cage.}
    \label{fig:NvDEx-field-cage}
\end{figure}

A cross-sectional view of a prototype field cage (FC) design is shown in Fig. \ref{fig:NvDEx-field-cage}. The FC is isolated from the inner copper shielding by a 20 mm-thick POM cylinder. The FC will be made of flexible printed circuit (FPC) sheets sized 315 mm$\times$423 mm. Each FPC has 5 mm-wide copper strips with a pitch of 6 mm on both sides. Three snap-off holes at both ends and the center of each copper strip will be used to align and fix the FPC sheets onto a 10 mm-thick POM supporting cylinder with screws. Two copper rings will be mounted at the two ends of the POM supporting cylinder. The copper strips and copper rings will be connected using low radioactive background resistors. The cathode of the TPC will be a low radioactive background copper plane mounted on the inner copper shielding disk, isolated with a POM layer of thickness 25 mm. 8 pogopins will be used to ensure a good connection between the cathode plane and the copper ring on the end of the cylinder part of the FC once the pressure chamber is closed.

\begin{figure}[ht]
    \centering
    \includegraphics[width=0.6\linewidth]{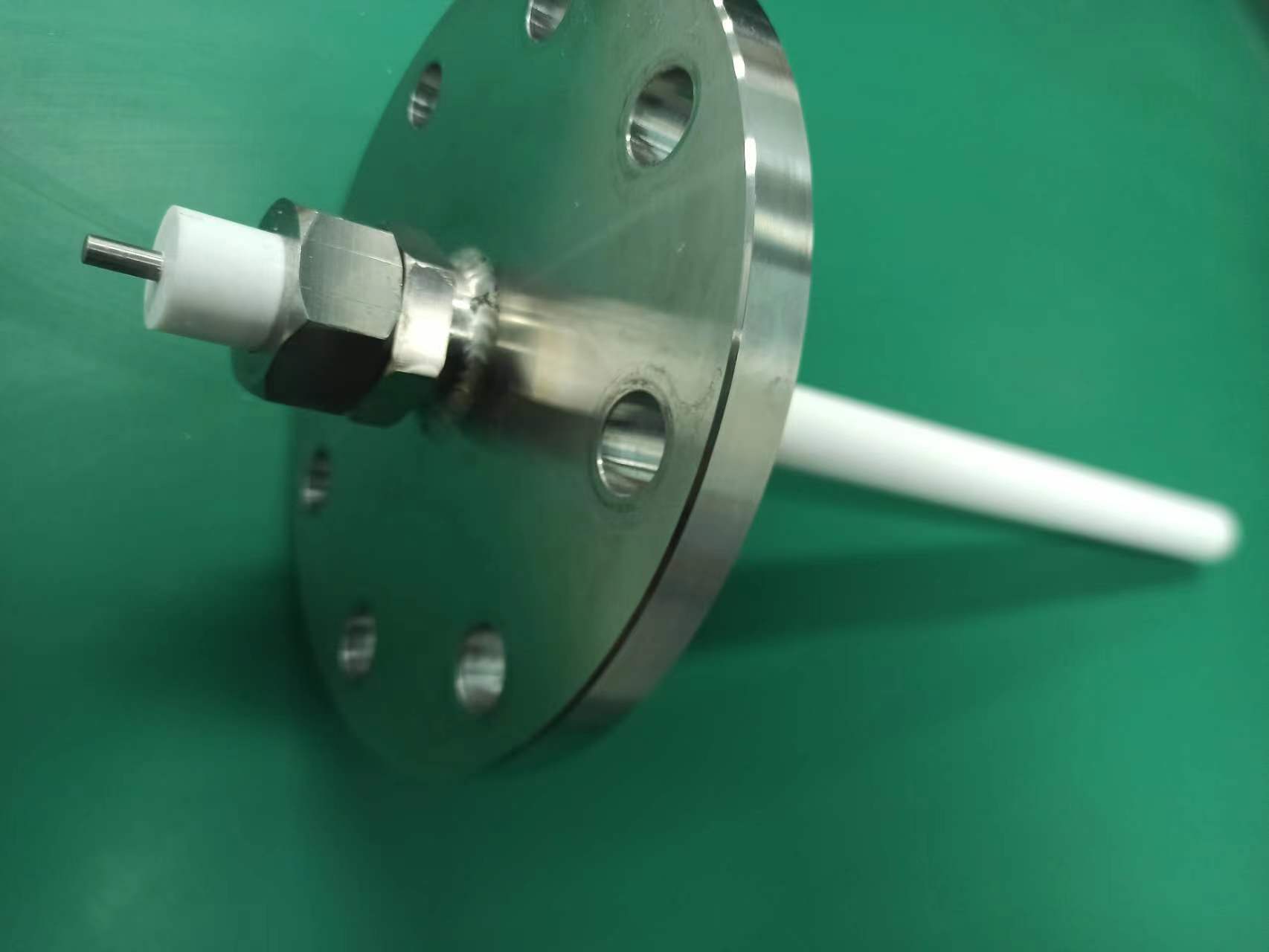}
    \caption{Image of the high voltage feedthrough.}
    \label{fig:NvDEx-filed-cage-HV-feedthrough}
\end{figure}

A high-voltage feedthrough will be connected to the cathode by spring pins. It is constructed using a compression seal approach, as shown in Fig. \ref{fig:NvDEx-filed-cage-HV-feedthrough}. A metal rod is pressed into a Polytetrafluoroethylene (PTEF) seal ring by clamping nuts on a DN80 flange. The feedthrough has been tested with a high voltage of 100 kV and for leak-tightness in nitrogen at 1.0 MPa.

\subsection{Topmetal-S Sensor and Readout Plane}
\label{sec:Topmetal}
\begin{figure}[ht]
    \centering
    \includegraphics[width=0.8\linewidth]{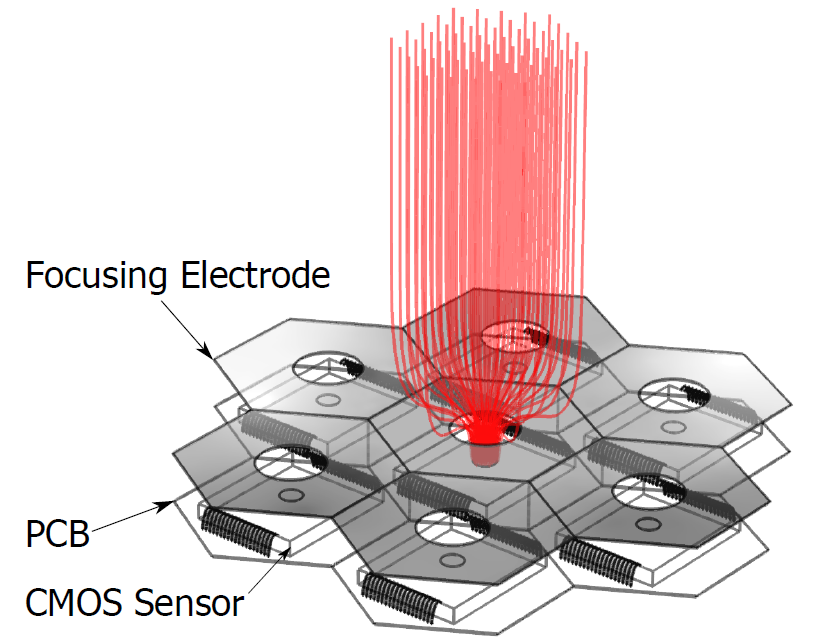}
    \caption{Topmetal-S sensors tiled in a hexagonal pattern to form a charge readout plane without gas gain.}
    \label{fig:TopmetalSplane}
\end{figure}

Around 10k CMOS sensors, named Topmetal-S, arranged in a hexagonal pattern as shown in Fig. \ref{fig:TopmetalSplane}, will be directly placed at the site of charge measurement to collect ionization charges without avalanche multiplication.
A perforated focusing electrode is placed above the readout plane with round holes aligned with the charge collection electrodes on the Topmetal-S sensors concentrically. The focusing structure ensures all charges eventually land on the charge collection electrode for maximum charge collection efficiency. 

Each Topmetal-S sensor is integrated with a charge collection electrode, a front-end amplifier, and data processing circuits. The charge collection electrode is an exposed hexagonal top-most metal with a diameter of about 1 mm. The charge signal collected on the Topmetal electrode is directly DC coupled to the charge sensitive pre-amplifier (CSA). The structure of the CSA in the prototype Topmetal-S sensor is a folded cascade amplifier with a feedback capacitor and a feedback transistor. The decay time of the CSA can be adjusted by changing the gate voltage of the feedback transistor.

\begin{figure}[ht]
    \centering
    \includegraphics[width=\linewidth]{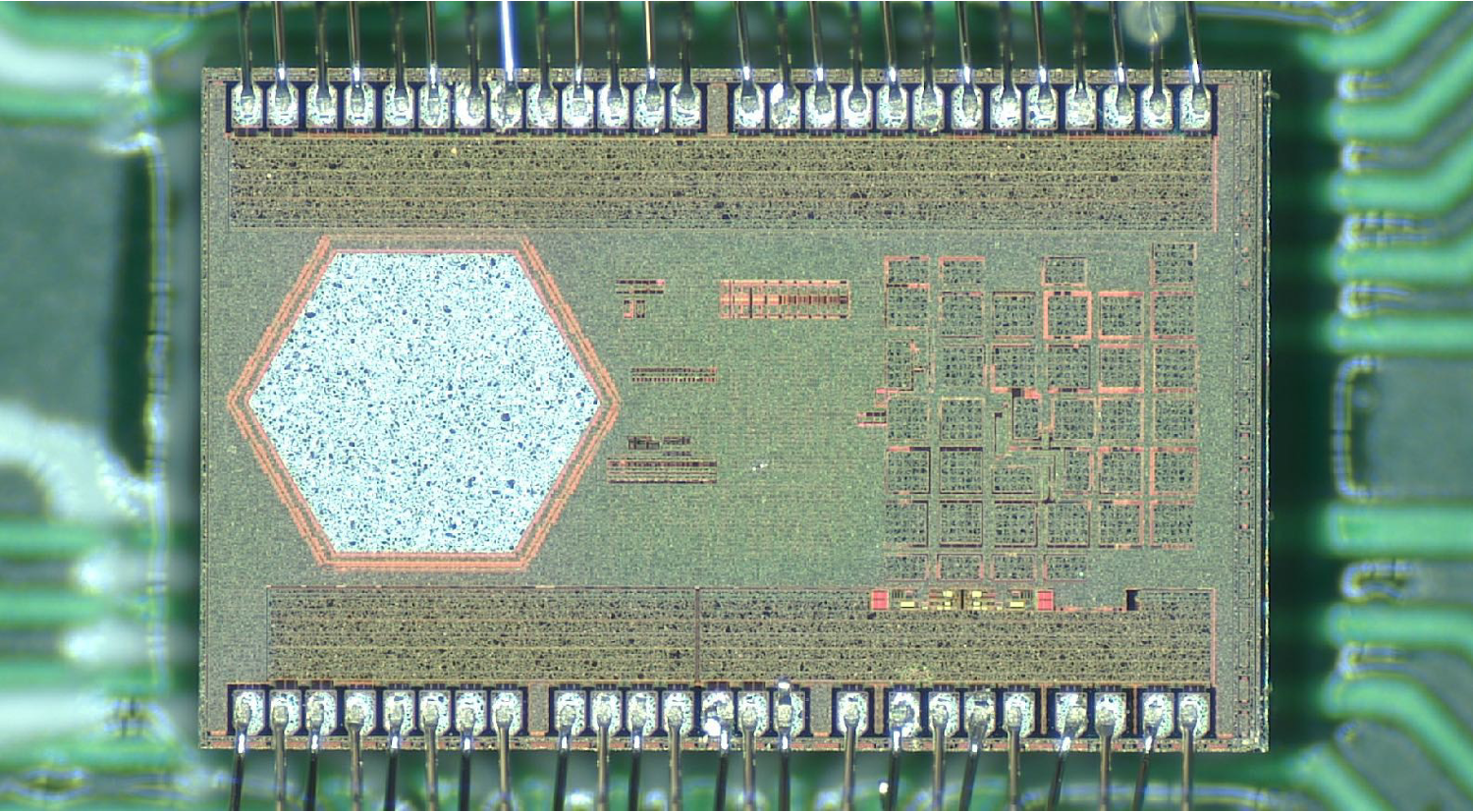}
    \caption{Photo of the Topmetal-S sensor chip.}
    \label{fig:Topmetal-S}
\end{figure}

Due to the stringent noise requirement, the analog signal of the CSA output must be digitized immediately. Thus, an in-chip Analog-to-Digital Converter (ADC) is designed to minimize the analog signal transfer. The ADC should have a noise floor well below the noise of the CSA of about 1 mV and a large enough dynamic range to cover the possible input charge range up to about 40 ke$^{-}$. A sigma-delta (SD) ADC is used in the Topmetal-S sensor. It comprises an SD modulator (analog  part) with coarse quantizers and a decimation filter (digital part) together to produce a data-stream output. By sampling the input signal at a frequency significantly higher than the signal bandwidth (oversampling), most of the noise shifts beyond the band of interest. A decimation filter further attenuates the out-of-band noise to achieve an improved signal-to-noise ratio. A photo of the Topmetal-S sensor chip is shown in Fig. \ref{fig:Topmetal-S}.

Since the sensors are densely packed on the plane, the number of available paths for the routing signal out of the plane is limited. Beyond a certain plane size or total number of sensors, routing every signal from all sensors out becomes impractical. Digitized data must be communicated through an inter-sensor network. Therefore, the circuitry that handles data processing and communication must be integrated into the sensor. A distributed, self-organizing, and fault-tolerant readout network is proposed with the Topmetal-S sensor. The proposed scheme forms a sensor network by establishing a local connection between the adjacent sensors. Each sensor integrates a router as a network node; hence, each sensor not only generates and transmits its data but also forwards the data from its adjacent nodes. Finally, the data are received by a data acquisition system, which is directly connected to the network edge and used to transmit the data between the sensor network and the computer.

\subsection{Data Acquisition}
\label{sec:DAQ}

\begin{figure}
    \centering
    \includegraphics[width=\linewidth]{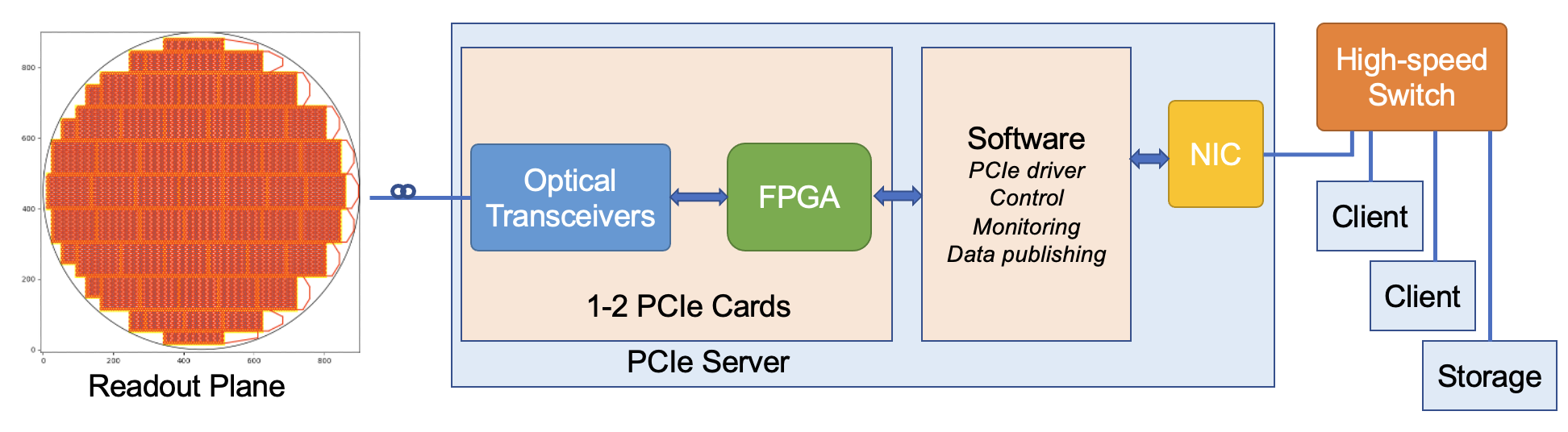}
    \caption{Architecture of the DAQ system.}
    \label{fig:NvDEx-daq}
\end{figure}

Based on the two-dimensional distributed network formed by the digital part of the Topmetal-S sensors, the digitized waveform of each CSA output will be transmitted to the edge of the plane as the streaming readout. The speed of the data chain could go up to 45 Mbps. As shown in Figure\,\ref{fig:NvDEx-daq}, the full plane is split into modules with different sizes to cover the end cap as much as possible. All the streaming data chains end in the modules on the right side, where the data are further encoded and aggregated into high-speed links with a speed of a few Gbps, by the commercial transceiver chips. Depending the on orientation of the sensors and how the sensors on the edge columns are connected to the transceiver, there will be 20$\sim$50 bidirectional high-speed links in total to connect the readout plane and the DAQ system in the back-end. In the other direction, the control data streams from the DAQ system are transmitted towards the left side of the readout plane. 

The flexible Printed Circuit Board (PCB) modules will be fabricated with radiopure material. CMOS chips, such as the Topmetal-S sensors, are known to be low in radioactive contamination. Other components, including the capacitors, resistors, transceivers, and some power chips, will be selected carefully. The radioactivity measurements will be done in the CJPL. Besides the material and components, any tools or materials used during the assembly procedures should also be clean enough.

A PCIe-based DAQ system will be built in the back-end to communicate with the front-end electronics on the readout plane via high-speed fiber optic links. A similar PCIe form factor has been adopted by dozens of large-scale experiments, such as the ATLAS experiment at the LHC and the sPHENIX experiment at the RHIC\,\cite{felix_2016,felix_2020}. The streaming data from all sensors are received and decoded by FPGAs on the PCIe cards. The data processing, event building, and filtering can be flexibly placed in the chain from the FPGA firmware to the software. The raw data and kinds of intermediate-stage data will be streamed from the DAQ server to a high-speed switch. Any client connected to the network can remotely subscribe to the data and implement further data analysis.

\subsection{External Shielding}
\label{sec:externalShielding}

\begin{figure}[ht]
    \centering
    \includegraphics[width=\linewidth]{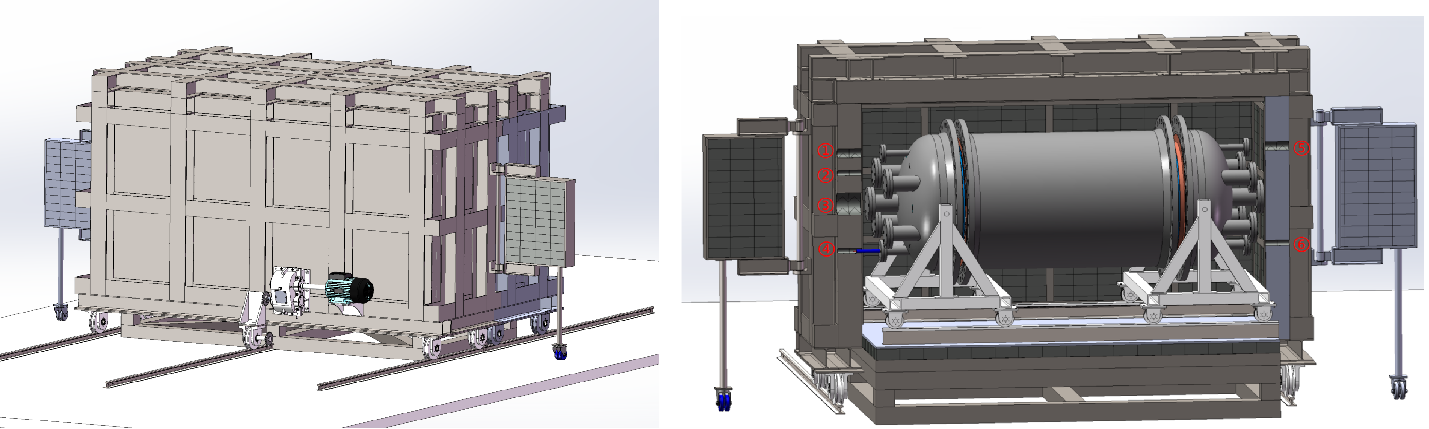}
    \caption{Design of the external shielding.}
    \label{fig:external_shielding}
\end{figure}

An external lead shielding with a thickness of 20 cm will be built outside the pressure chamber to protect the NvDEx detector from environmental radiation. The preliminary design of the external lead shielding is shown in Fig.\ref{fig:external_shielding}. It comprises two mobile halves and a fixed base on which the pressure chamber sits. The mobile halves, including the side walls and the top, are installed on a mobile base connected to a transmission system so that they can be adjusted to enable opening and operations on the pressure chamber.
The fixed base is installed on a vibration isolation system to minimize the influence of vibration on the experimental measurements.
The vacuum and gas pipes, high- and low-voltage cables, and the optical fibers will pass the external lead shielding through several holes at the joints of the two mobile halves. Two shielding doors will be installed to prevent radiation from going through these holes directly.
The lead bricks and the steel structure inside the lead layer will be selected and tested for radioactive contamination.

Lead is very effective for stopping external $\gamma$ radiation; however, it is  not a good shielding material for neutrons. High-density polyethylene (HDPE) blocks will be placed as much as possible between the pressure chamber and external lead shielding to slow down and absorb neutrons. The lead shielding exterior will also be covered with a 30 cm-thick layer of HDPE. The HDPE material will also be tested for radioactivity.

\subsection{Gas System}
\label{sec:gasSystem}

The working medium of \NvDEx \ is highly toxic \SeF6 \ gas. When there is moisture in it, \SeF6 \ can easily decompose and produce corrosive HF, which may damage detector components and/or cause toxic gas leakage. 
The gas system of \NvDEx \ can fill the pressure chamber with \SeF6 \ to the working pressure of 1 MPa, discharge the gas from the pressure chamber, and safely store it during experimental maintenance.
During the lifetime of the experiment, the pressure vessel will be pressurized, depressurized, and vacuumized many times, and the pressure vessel and part of the gas system will operate at a pressure of 1 MPa for years during data collection. Thus, gas tightness and reliability are critical for the \NvDEx \ gas system.

\begin{figure*}[htbp]
    \centering
    \includegraphics[width=0.8\textwidth]{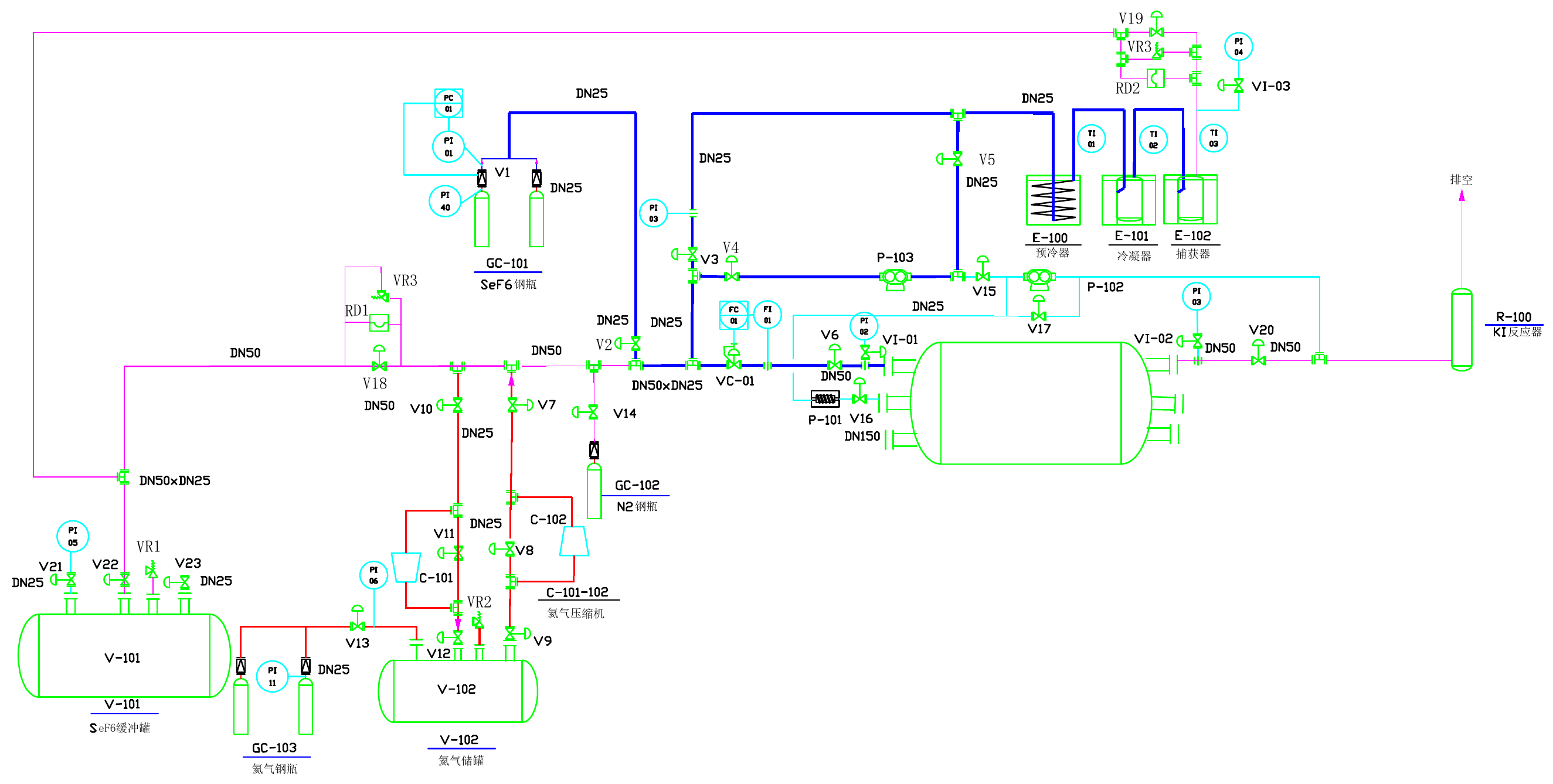}
    \caption{Schematic of the \NvDEx \ gas system.}
    \label{fig:gas_diagram}
\end{figure*}

With these considerations, the schematic of the \NvDEx \ gas system is designed as shown in Fig. \ref{fig:gas_diagram}.

The pressure chamber is connected to a turbomolecular vacuum pump and a dry vacuum pump, which can vacuumize the chamber and the gas system before filling \SeF6 \ gas. This can minimize contamination of the \SeF6 \ gas by air, moisture, and radon to avoid corrosion and radiation background from the gas.

Every time before filling the toxic \SeF6 \ gas to the chamber, \SF6, which has similar properties as \SeF6 \ but is non-toxic, is filled into the system with a pressure of 1 MPa to test the gas tightness of the pressure chamber and gas system.
After the test, \SF6 \ is compressed into a \SF6 \ storage tank for future use. 

In emergencies like gas leakage, fire, an earthquake, a power outage, etc., \SeF6 \ gas in the system will be released into an emergency pressure relief tank within 10 seconds so that the pressure in the system is below atmospheric pressure to ensure personnel and environmental safety.

Before experimental maintenance, \SeF6 \ gas is discharged from the pressure chamber and condensed by the low temperature in a precooler and two condensers. After the \SeF6 \ saturated vapor pressure is reached, a dry vacuum pump pumps the gas from the pressure chamber to the condensers. Then, any trace amount of \SeF6 \ left in the system is flushed by nitrogen gas into a 
potassium iodide (KI) reactor to be absorbed. The \SeF6 is safely stored in the condensers at low a temperature as solid and vapor below atmospheric pressure, which can be refilled in the pressure chamber in the future.

All the components for the gas system have been purchased and are waiting to be assembled once the experimental pressure chamber manufacture is complete. Before moving to CJPL, the system will be commissioned and tested in the above-ground lab with \SF6 \ gas.

\subsection{Airtight Clean Room}
\label{sec:cleanRoom}

The entire experimental setup is placed in an airtight clean room to ensure the whole experimental process is safe.
Enough potassium iodide (KI) reagent is placed in the airtight clean room as a second line of defense for environmental safety.

The control room is located outside the airtight clean room.
Under normal working conditions, the airtight clean room is in an airtight state, and the entire experimental setup, including the gas system, can be controlled remotely.
In case of a gas leakage, the gas system releases the \SeF6~ gas into the pressure relief tank, and the KI reagent placed in the airtight clean room reacts with the leaked \SeF6~ and absorbs it.
The airtight clean room is divided into three areas that are airtightly isolated from each other.
The main experimental setup and most gas systems are located in one area. Two \SeF6~ condensers that serve as backups for each other are placed in the other two areas.
When people need to operate the experiment in the main area of the experimental setup, all \SeF6~ gas should be condensed in one of the condensers first to eliminate \SeF6~ gas in the main area. Further, the pressure inside the condenser is the vapor pressure of \SeF6, which is far below atmospheric pressure, so the leakage of a large amount of gas is impossible.
In rare cases where people need to conduct on-site operations on a condenser, \SeF6~ is first condensed in the other condenser.
Therefore, people are always in an airtight clean room without \SeF6~ gas.

The airtight clean room needs to reach a clean level of 100k to control the background radioactivity by preventing the surface of the detector from being contaminated.
The temperature in the airtight clean room needs to be controlled at 22$\pm$2$^{\circ}$C to minimize the gas convection in the TPC.
Concurrently, the humidity must be 30$\pm$10\% to avoid moisture condensation at cold spots in the experimental setup, such as the cooling plate on the pressure chamber.

\section{Background and Sensitivity Estimation}
\label{sec:simulation}

\subsection{Natural Radioactive $\gamma$ Background}
\label{sec:gammaBackground}

Radioactive isotopes are always present in the walls of the hall and in the rock surrounding them, as well as in the materials of the experimental setup. The decay of these isotopes produces a large amount of $\alpha$, $\beta$, and $\gamma$ particles. 
However, the first two will mostly be stopped without causing a detectable background unless they are generated in or near the sensitive volume. So, we mainly focus on the latter. $\gamma$ particles penetrating into the sensitive volume may interact with the gas and generate free electrons visible to the detector, making them a major background source for \NvDEx.

A 20 cm-thick external lead shielding and a 12 cm-thick inner copper shielding outside and in the pressure chamber shield the external $\gamma$ particles. The thicknesses of the shielding layers are optimized using Geant4 simulations \cite{Allison:2006ve,GEANT4:2002zbu,Allison:2016lfl}. Owing to the presence of radioactive contamination in the shielding materials, the $\gamma$ flux in the sensitive volume does not always decrease as the shielding thickness increases.

Assuming Topmetal-S sensors can achieve 30 e$^-$ noise level, on average, 25 sensors are fired per \0vbb event, and the ionization energy and Fano factor of \SeF6 are equal to that of \SF6 (32 eV, 0.19), then the FWHM energy resolution calculated is 0.51\%. Considering deviations from the above assumptions and imperfections of the experiment, to be conservative, we assume 1\% FWHM energy resolution and use this energy range (2.98-3.01 MeV) as the region of interest (ROI) in the following background estimations.

Most of the ROI $\gamma$ background comes from the decay of $^{214}$Bi from the $^{238}$U decay chain. The probability of having a $\gamma$ with energy larger than 2.9 MeV from a $^{214}$Bi decay is $6.8\times10^{-4}$ \cite{Brown20181short}, which is quite rare. This is the advantage of $^{82}$Se's high $Q_{\beta\beta}$ for \NvDEx.
$^{208}$Tl from the $^{232}$Th decay chain contributes about one order of magnitude less ROI background than $^{214}$Bi, with only around $8.5\times10^{-5}$ \cite{Brown20181short} of the produced $\gamma$ having energy around and above the ROI. Thus, we only focus on the $^{238}$U decay chain in our $\gamma$ background simulation and estimation.

The $^{238}$U radiation activities of materials considered in our simulations are listed in table \ref{tab:Contamination}.
The $^{238}$U activity of the concrete in the experimental hall walls is from a measurement at CJPL \cite{Ma:2020rpd}. The contribution of the rocks is negligible since the contamination rate of the rocks is significantly lower than that of  the concrete. For other materials in the experimental setup, we used the measurements from the NEXT experiment \cite{NEXT:2012zwy}. These numbers will be updated using measurements of materials for \NvDEx \ in the future. The activity of the POM material used for the TPC field cage is assumed to be identical to that of HDPE in NEXT. Many other parts or materials, for example, the \SeF6 gas, Topmetal-S sensors, flexible PCBs used in the field cage and readout plane, bolts on the pressure chamber, and various steel supporting structures outside the pressure chamber, are not considered in the current simulation because of their relatively smaller sizes, lower background contributions, complications in building the geometry model, and/or lack of knowledge on their radioactive activities. For these reasons, the current background simulation only serves as a rough estimation and guides hardware developments.

\begin{table}[htbp]
\centering
\caption{\label{tab:Contamination} Activity assumed for each material}
\smallskip
\begin{tabular}{lll}
\hline
Material & Subsystem & $^{238}$U Activity (mBq/kg) \\
\hline
Concrete & Experimental hall & $6.8\times10^3$ \cite{Ma:2020rpd}\\
Lead & External shielding & 0.37 \cite{NEXT:2012zwy}\\
HDPE & External shielding & 0.23 \cite{NEXT:2012zwy}\\
Steel & Pressure vessel & 1.9 \cite{NEXT:2012zwy}\\
Copper & Inner copper shielding & 0.012\cite{NEXT:2012zwy}\\
POM & Field cage & 0.23\cite{NEXT:2012zwy}\\
\hline
\end{tabular}
\end{table}

The natural radioactive $\gamma$ background simulation results are reported in Table \ref{tab:BackgroundGamma}. The largest contribution comes from the POM of the field cage because it is close to the sensitive gas volume. The total ROI $\gamma$ background level is about 0.4 evts/yr. Note that, like most other types of backgrounds described in the following sub-sections, this background can be further suppressed by one order of magnitude using a neural network considering event topology information ~\cite{Nygren:2018ewr}, to the order of only 0.04 evts/yr in the ROI.

\begin{table}[htbp]
\centering
\caption{\label{tab:BackgroundGamma} $\gamma$ background from different sources without suppression using event topology}
\smallskip
\begin{tabular}{llll}
\hline
\multicolumn{2}{c}{Source} & \multicolumn{2}{c}{Background in ROI} \\
Material & Subsystem & evts/yr & $10^{-5}$evts/(keV kg yr)\\
\hline
Concrete & Experimental hall & 0.004 & 0.12 \\
Lead & External shielding & 0.003 & 0.09 \\
HDPE & External shielding & 0.005 & 0.16 \\
Steel & Pressure vessel & 0.026 & 0.86 \\
Copper & Inner copper shielding & 0.050 & 1.67 \\
POM & Field cage & 0.330 & 10.99 \\
\hline
Total & & 0.42 & 13.9 \\
\hline
\end{tabular}
\end{table}

\subsection{Neutron Background}
\label{sec:neutronBackground}

Radioactive decay can also emit neutrons. These events are quite rare; however, since it is significantly more difficult to stop neutrons than $\gamma$, the former can arrive more easily at the sensitive volume. Neutrons do not create ionization signals directly, but they can activate nuclei inside the detector, creating $\gamma$ via (n,$\gamma$) and (n,n'$\gamma$) reactions.

Neutrons can also create unstable isotopes, which decay and emit $\alpha$ or $\beta$ particles. However, unlike $\gamma$ particles, these $\alpha$ and $\beta$ particles can generate background only if they are created in or near the fiducial volume; otherwise, they would be stopped almost immediately.

Among the isotopes that can be created in the fiducial volume by the neutrons, four contribute to the background: $^{20}$F, $^{16}$N, $^{19}$O, and $^{83}$Se. They can be produced through the following reactions.
\begin{eqnarray}
    ^{19}F+n\rightarrow ^{20}F \nonumber \\
    ^{19}F+n\rightarrow ^{19}O+p \nonumber \\
    ^{19}F+n\rightarrow ^{16}N+\alpha \nonumber \\
    ^{82}Se +n \rightarrow ^{83}Se
\end{eqnarray}
In principle, other unstable isotopes can also be created, but their Q-values are significantly lower than that of the $^{82}$Se \0vbb signal; therefore, we do not need to consider them.

We studied the neutron-induced background rate using the fast neutron spectrum measured at CJPL and reported in \cite{Hu:2016vbu} using Geant4 and FLUKA \cite{Bohlen:2014buj,Ferrari:2005zk,Battistoni:2015epi,Ahdida:2022gjl} packages. The results can be found in~\cite{wang2023neutron}.

Neutron-induced $\gamma$ is the main source of the neutron background, while the $\alpha$ and $\beta$ decays of unstable isotopes created by neutrons are strongly subdominant. Without any HDPE shielding, the ROI background rate of the former will be 1492 evts/yr, while that of the latter will be 39 evts/yr.

Thus, HDPE shielding is added to stop neutrons. Unlike high-Z materials such as copper and lead, HDPE contains many hydrogen nuclei, which can slow down and absorb neutrons very effectively.
HDPE blocks are placed between the external lead shielding and pressure vessel, and a 30 cm-thick HDPE layer is placed outside the Pb shielding. Thus, it is possible to reduce the neutron-induced background to 0.03 evts/yr, strongly subdominant with respect to the $\gamma$ background directly from natural radioactivity.

\subsection{Cosmogenic Background}
\label{sec:cosmogenicBackground}
When materials used in the experiment are exposed to cosmic rays during production and transportation on the ground, they are activated, creating relatively long-lived isotopes. 
The half-lives of these isotopes, although much shorter than those of $^{238}$U and $^{232}$Th described in Sub-section \ref{sec:gammaBackground}, are long enough (for example, months to years) to make them an important background source, even after the materials are placed underground.

The spallation process by high-energy cosmic nucleons is a dominant process in the cosmogenic production of radionuclides. However, other reactions such as capture can also be important in some cases. 
Spallation reactions can produce many radionuclides, depending on the atomic number of the target material. On the Earth’s surface, isotope production is dominated by neutrons because protons are absorbed by the atmosphere.

Cosmogenic activation can be minimized by reducing surface exposure, for example, by shielding against cosmic rays, avoiding flights and storing on the ground, or even producing materials underground. Purification techniques can also eliminate many of the induced isotopes. However, these preventive measures make the experiment preparation more complex. Consequently, it is advisable to assess the relevance of the material exposure to cosmic rays for the experiments and its effect on sensitivity. To quantify the induced activity, A, of an isotope with decay constant $\lambda$, the production rate R of the isotope in the considered target and the exposure history must be well-known. Specifically, A can be computed as:
\begin{equation}
    A=R(1-e^{-t_{exp}/\lambda})e^{-t_{cool}/\lambda}
\end{equation}
 where $t_{exp}$ is the time of exposure to cosmic rays and $t_{cool}$ is the cooling time (time spent underground once shielded from cosmic rays).

Some direct measurements of production rates have been conducted for a few materials exposed in well-controlled conditions. However, in many cases, production rates must be evaluated from the flux (per unit energy) of cosmic rays, $\phi$, and the isotope production cross-section, $\sigma$, both depending on the particle energy E:

\begin{equation}
    R=N_{t}\int \sigma(E) \phi(E) dE   ,
\end{equation}
where $N_{t}$ denotes the number of target nuclei. We used the ACTIVIA code \cite{ACTIVIA} to calculate the cosmogenic activation of the various materials used in \NvDEx. The cosmogenic activation rate of various radioisotopes in \SeF6 \ gas, copper, lead, and steel and activities after exposure and cooling for certain durations are shown in Tables \ref{tab:cosmicGenSe}, \ref{tab:cosmicGenCu}, \ref{tab:cosmicGenPb}, and \ref{tab:cosmicGenSteel}. Only isotopes with relatively long half-lives and high Q-values are listed because other isotopes do not create a background in the 0$\nu\beta\beta$ decay ROI after being placed underground to cool for some time.

\begin{table}[h!]
    \centering
    \caption{Cosmogenic activation rate of various radioisotopes in enriched \Se82 and activities after exposure at sea level and cooling for certain durations.}
    \begin{tabular}{llllllll}
    \hline\hline
    Isotope & Q-value & Half-life & & \multicolumn{2}{c}{Production rate} & Activity & Activity \\
    & (keV) & (d)  & & \multicolumn{2}{c}{(atoms/kg/d)} & after 2yr & after 1yr \\ 
    &&&& Calc. & Expt. & exposure & cooling \\
    &&&&&& ($\mu$Bq/kg) & ($\mu$Bq/kg) \\

    \hline
      $^{54}$Mn & 1377 & 312 && 0.37 &- & 3.4 & 1.5  \\
      $^{56}$Co & 4566 & 77.3 && 0.04 & - & 0.46 & 0.02  \\
      $^{57}$Co & 836 & 272 && 0.14 & - & 1.4 & 0.54  \\
      $^{58}$Co & 2307  & 70.9  && 0.83 & - & 9.6 & 0.27  \\
      $^{60}$Co & 2824 & 1.92$\times$10$^{3}$ && 0.11 & - & 0.29 & 0.26  \\
      $^{75}$Se & 864 & 120 && 14.9 & - & 170 & 20.6  \\
    \hline\hline
    \end{tabular}

    \label{tab:cosmicGenSe}
\end{table}

From Tab. \ref{tab:cosmicGenSe}, we can see that $^{56}$Co is the most important cosmogenic background isotope from \Se82~. It has a Q-value of 4566 keV, which is above the \Se82 \ Q-value, and a half-life of 77.3 days, long enough to have considerable activity of 0.02 $\mu$Bq/kg after two years of exposure at sea level and one year of cooling time. Cobalt fluorides are solid rather than gas at room temperature, and so far, we do not know whether or how much of the generated single-molecular $^{56}$Co fluorides will remain in the gas after \SeF6 production, storage,  and transportation. Assuming conservatively that all the $^{56}$Co stays in the gas and reaches the sensitive volume of the experiment, for 100kg of \82SeF6 with $^{56}$Co activity of 0.02 $\mu$Bq/kg, approximately 26 decays will occur per year. Thus, energy deposition in the ROI will be minimal.

Cosmogenic isotopes from $^{19}$F, with mass numbers no larger than 19, do not have relatively large Q-values and long lifetimes simultaneously and thus will not constitute important cosmogenic background contributions.

\begin{table}[h!]
    \centering
    \caption{Cosmogenic activation rate of various radioisotopes in copper and activities after exposure at sea level and cooling for certain durations.}
    \begin{tabular}{llllllll}
    \hline\hline
    Isotope & Q-value & Half-life & & \multicolumn{2}{c}{Production rate} & Activity & Activity \\
    & (keV) & (d)  & & \multicolumn{2}{c}{(atoms/kg/d)} & after 2yr & after 1yr \\ 
    &&&& Calc. & Expt.~\cite{LAUBENSTEIN2009750} & exposure & cooling \\
    &&&&&& ($\mu$Bq/kg) & ($\mu$Bq/kg) \\

    \hline
      $^{46}$Sc & 2367 & 83.8 && 3.1 & 2.18$\pm$0.74 & 36 & 1.7  \\
      $^{54}$Mn & 1377 & 312 && 14.3 & 8.85$\pm$0.86 & 133 & 59  \\
      $^{59}$Fe & 1565 & 44.5 && 4.2 & 18.7$\pm$4.9 & 49 & 0.2  \\
      $^{56}$Co & 4566 & 77.3 && 8.7 & 9.5$\pm$1.2 & 101 & 3.8  \\
      $^{57}$Co & 836 & 272 && 32.5 & 74$\pm$17 & 318 & 125  \\
      $^{58}$Co & 2307  & 70.9  && 56.6 & 67.9$\pm$3.7 & 655 & 18  \\
      $^{60}$Co & 2824 & 1.92$\times$10$^{3}$ && 26.3 & 86.4$\pm$7.8 & 71 & 62  \\
    \hline\hline
    \end{tabular}

    \label{tab:cosmicGenCu}
\end{table}

The most important cosmogenic background isotope in copper is also $^{56}$Co, as shown in Tab. \ref{tab:cosmicGenCu}. Considering that the inner copper shielding will be assembled and tested with the pressure chamber at the Institute of Modern Physics in Lanzhou, with an altitude of about 1500m, the exposed cosmic neutron flux is about 3.2 times higher than at sea level. The $^{56}$Co activity will be about 323 $\mu$Bq/kg after 2 years of exposure.
Using the simulation framework described in Sub-section \ref{sec:gammaBackground}, we find that the ROI background in the sensitive volume from $^{56}$Co emitted $\gamma$'s is as high as about 3700 evts/yr, which is much higher than the natural radiation $\gamma$ background. After three years of cooling, the ROI background will drop to a subdominant level of 0.19 evts/yr. So, it is important to place the inner copper shielding underground for cooling as early as possible.

The other isotopes in Table \ref{tab:cosmicGenCu} have Q-values lower than those of \Se82. Therefore, they do not contribute to the ROI background alone. However, because the drift velocity of ions in \NvDEx \ TPC is slow, there is a chance that ionization due to the $\gamma$'s from these isotopes adds up with other background sources, forming the so-called pile-up event backgrounds and reaching the \Se82 Q-value. This is described in Subsection \ref{sec:pileUpBackground}, taking $^{60}$Co, which has a relatively long half-life and high Q-value, as an example of cosmogenic background isotopes.

\begin{table}[h!]
    \centering
    \caption{Cosmogenic activation rate of various radioisotopes in lead and activities after exposure at sea level and cooling for certain durations. For short-lived isotopes with very long-lived parents (given in parentheses), we have considered the half-life of parent isotopes.}
    \begin{tabular}{llllllll}
    \hline\hline
    Isotope & Q- & Half-life & & \multicolumn{2}{c}{Production} & Activity & Activity \\
    & value & (d) & & \multicolumn{2}{c}{rate} & after 2yr & after 1yr \\ 
    & (keV) &&& \multicolumn{2}{c}{(atoms/kg/d)} & exposure & cooling \\
    &&&& Calc. & Expt. & ($\mu$Bq/kg) & ($\mu$Bq/kg) \\

    \hline
      $^{56}$Co & 4566 & 77.3 && 0.026 & -- & 0.30 & 0.01  \\
      $^{57}$Co & 836 & 272 && 0.047 & -- & 0.46 & 0.18  \\
      $^{58}$Co & 2307 & 70.9 && 0.127 & -- & 1.47 & 0.04  \\
      $^{60}$Co & 2824 & 1.92$\times$10$^{3}$ && 0.008 & -- & 0.02 & 0.02  \\
      $^{194}$Au ($^{194}$Hg) & 2501 & 1.90$\times$10$^{5}$ && 5.52 & -- & 0.17 & 0.17  \\
      $^{202}$Tl($^{202}$Pb) & 2398 & 1.93$\times$10$^{7}$ && 120 & -- & 0.04 & 0.04  \\
      $^{207}$Bi & 1363 & 1.15$\times$10$^{4}$  && 1.42 & -- & 0.71 & 0.69  \\
    \hline\hline
    \end{tabular}

    \label{tab:cosmicGenPb}
\end{table}

\begin{table}[h!]
    \centering
    \caption{Cosmogenic activation rate of various radioisotopes in steel and activities after exposure at sea level and cooling for certain durations.}
    \begin{tabular}{llllllll}
    \hline\hline
    Isotope & Q-value & Half-life & & \multicolumn{2}{c}{Production rate} & Activity & Activity \\
    & (keV) & (d)  & & \multicolumn{2}{c}{(atoms/kg/d)} & after 2yr & after 1yr \\ 
    &&&& Calc. & Expt. & exposure & cooling \\
    &&&&&& ($\mu$Bq/kg) & ($\mu$Bq/kg) \\

    \hline
    $^{48}$V & 4012 & 16.0 && 21.6 & -- & 250 & 3.4$\times$10$^{-5}$  \\  
    $^{52}$Mn & 4712 & 5.59 && 40.0 & -- & 463 & 1.0$\times$10$^{-17}$  \\
    $^{56}$Co & 4566 & 77.3 && 46.1 & -- & 533 & 20  \\
    $^{58}$Co & 2307 & 70.9 && 5.1 & -- & 59 & 1.7  \\
    $^{60}$Co & 2824 & 1.92$\times$10$^{3}$ && 0.24 & -- & 0.64 & 0.56  \\
    $^{88}$Y & 3623 & 107 && 0.042 & -- & 0.48 & 0.045  \\
    \hline\hline
    \end{tabular}

    \label{tab:cosmicGenSteel}
\end{table}

As shown in Tables \ref{tab:cosmicGenPb} and \ref{tab:cosmicGenSteel}, the production rates of cosmogenic backgrounds in lead and steel are either lower than or comparable to those in copper. Considering that the inner copper shielding shields them from the sensitive volume , their background contribution should be less important than the cosmogenic backgrounds in copper.

\subsection{Other Backgrounds}
\label{sec:backgroundsInGeneral}

We also considered the following background categories for \NvDEx, which are much lower than the natural radioactive $\gamma$, neutron, and cosmogenic backgrounds and thus can be neglected:
\begin{itemize}
    \item {\bf Natural radioactive $\alpha$ and $\beta$ background}: $\alpha$ and $\beta$ from natural radioactive isotopes have much shorter path lengths than $\gamma$ and neutrons. So, they can influence the experiment measurement only if the radioactive isotopes are in the gas or on the surface of the sensitive volume. For the \SeF6 gas, since both the Se and F$_{2}$ gas materials are obtainable with high purity (99.999\% for Se and 99.99\% for F$_{2}$), the radioactive isotope contamination should be low. This needs to be further studied and confirmed using ICP-MS measurements of the Se material or future analysis of the data from \NvDEx .\  Fluorides of U and Th are solids at room temperature. So far, we do not know whether or how much of the U and Th will stay in the gas and enter the sensitive volume after production, storage, and filling of \SeF6. Radioactive isotope contamination on the surface of the sensitive volume can be limited by carefully cleaning the components in the pressure chamber and gas system, especially directly on the surface of the sensitive volume, i.e., the inner surface of the field cage, the high voltage plate, and the focusing plane. When analyzing the data, $\alpha$ and $\beta$ from radioactive contamination on the surface of the sensitive volume can also be reduced by cutting the volume within a certain distance to the surface since the path length is within a few cm in the gas at 1 MPa.
    \item {\bf Radon background}: Radon, as a radioactive gas that can be emitted from radioactive isotopes in the underground environment and materials in the experiment setup, may become a background source if it gets into the sensitive volume of the experiment. Fresh air will be flushed through the clean room where the \NvDEx \ experiment is located to limit the level of radon  contamination during the assembly, operation, and maintenance of the experiment. Additionally, during the experiment, the pressure chamber is airtight with a gas pressure of 1 MPa inside. Thus, the amount of radon penetrating the chamber should be minimal. Radon can also be emitted from the materials inside the pressure chamber. Both $^{222}$Rn (half-life of 3.8 d) from the $^{238}$U chain and $^{220}$Rn (half-life of 55 s) from the $^{232}$Th chain undergo alpha decay into polonium. A portion of the products in the following decay chain may be negative or positive ions, which drift toward the anode or cathode in the electric field of the TPC.
    The $\alpha$ and $\beta$ from the decay of radon and its decay products (if they are not ions and drift to the anode or cathode) are further suppressed by the event topology characteristics in the TPC. This is because $\alpha$ forms thicker and straighter tracks than $\beta$, and a single $\beta$ event has one Bragg peak instead of two.
    $^{214}$Bi and $^{210}$Tl in the $^{222}$Rn decay chain have $\beta$ decay endpoint energy above the ROI and thus could contribute to the ROI background. However, $^{214}$Bi $\beta$ decay is immediately followed by a $^{214}$Po $\alpha$ decay with a half-life of 164 $\mu$s and $\alpha$ energy above 6 MeV. Thus, the sequential $\beta$ and $\alpha$ decays are reconstructed as one event, whose total energy will be far above the \Se82 $Q_{\beta\beta}$. The $^{210}$Tl, on the other hand, will only be generated from the $^{214}$Bi $\alpha$ decay with a small branching fraction of 0.02\%. So the ROI backgrounds from the $^{214}$Bi and $^{210}$Tl $\beta$ decays are limited.
    Other $\alpha$ and $\beta$ energies in the $^{222}$Rn and $^{220}$Rn decay chains are mostly away from the \Se82 $Q_{\beta\beta}$, further limiting the chance to create any background in the ROI. 
    $\gamma$ from the decay of radon and its decay products have a small chance of interacting with the gas, making a much smaller background contribution in ROI than the natural radioactive $\gamma$ directly from the materials and environment mentioned in Sub-section \ref{sec:gammaBackground}.
    \item {\bf Cosmic muon background}: As the deepest underground lab in the world, CJPL has a cosmic muon flux as low as $3.53 \pm 0.22 \textrm{ (stat.)} \pm 0.07 \textrm{ (sys.)}\times 10^{-10} \mathrm{cm}^{-2}\mathrm{s}^{-1}$~\cite{JNE:2020bwn}. \NvDEx-100, as a meter-scale experiment at CJPL, only observes a level of 0.3 cosmic muons per day. These muon background events show straight tracks through the TPC, can easily be distinguished from \0vbb signal events, and thus can be neglected. A very small fraction of muons interact with the gas, creating some radioactive isotopes. The chance of this kind of background falling into the energy ROI is also negligible, considering the low muon flux at CJPL. 
    \item {\bf Neutrino scattering background}: Neutrinos from various sources can easily penetrate into the sensitive volume of the detector, but the chance of a neutrino interacting with the gas is very low. The most important contribution comes from solar neutrinos scattering with electrons and yielding backgrounds at a rate lower than the order of 0.1 evts/ROI/ton/yr~\cite{Zhao:2016brs}. These single electron background events are further suppressed with event topology analysis, as mentioned in Sub-section \ref{sec:gammaBackground}.
    \item {\bf \2vbb decay background}: \2vbb decay events have exactly the same characteristics as \0vbb events except for their lower energy. As shown in Fig. \ref{fig:pileUpBackgroundNatural}, \2vbb decay only contributes $1\times10^{-6}$ background evts/yr based on the expected 1\% FWHM energy resolution of \NvDEx, which means it is not a major background source.
\end{itemize}

\subsection{Pile-up Event Background}
\label{sec:pileUpBackground}

\NvDEx-100 uses a TPC as the main detector.
The drift time for the approximately 160 cm maximum drift length is about 7 seconds. 
If one of the background events listed above happens, while the ionization charges of that event drift, another background event could happen near the cloud of the drifting charges from the first event. These two events could "pile up" on top of each other, arriving at the read-out plane at the same time and location, looking like a single event with energy equal to the total energy of these two events.
Thus, pile-up events tend to have higher energy than single background events.
Since all background rates drop dramatically with increasing energy, pile-up events have a higher chance of falling into the ROI than single background events.

\begin{figure}[ht]
    \centering
    \includegraphics[width=0.8\linewidth]{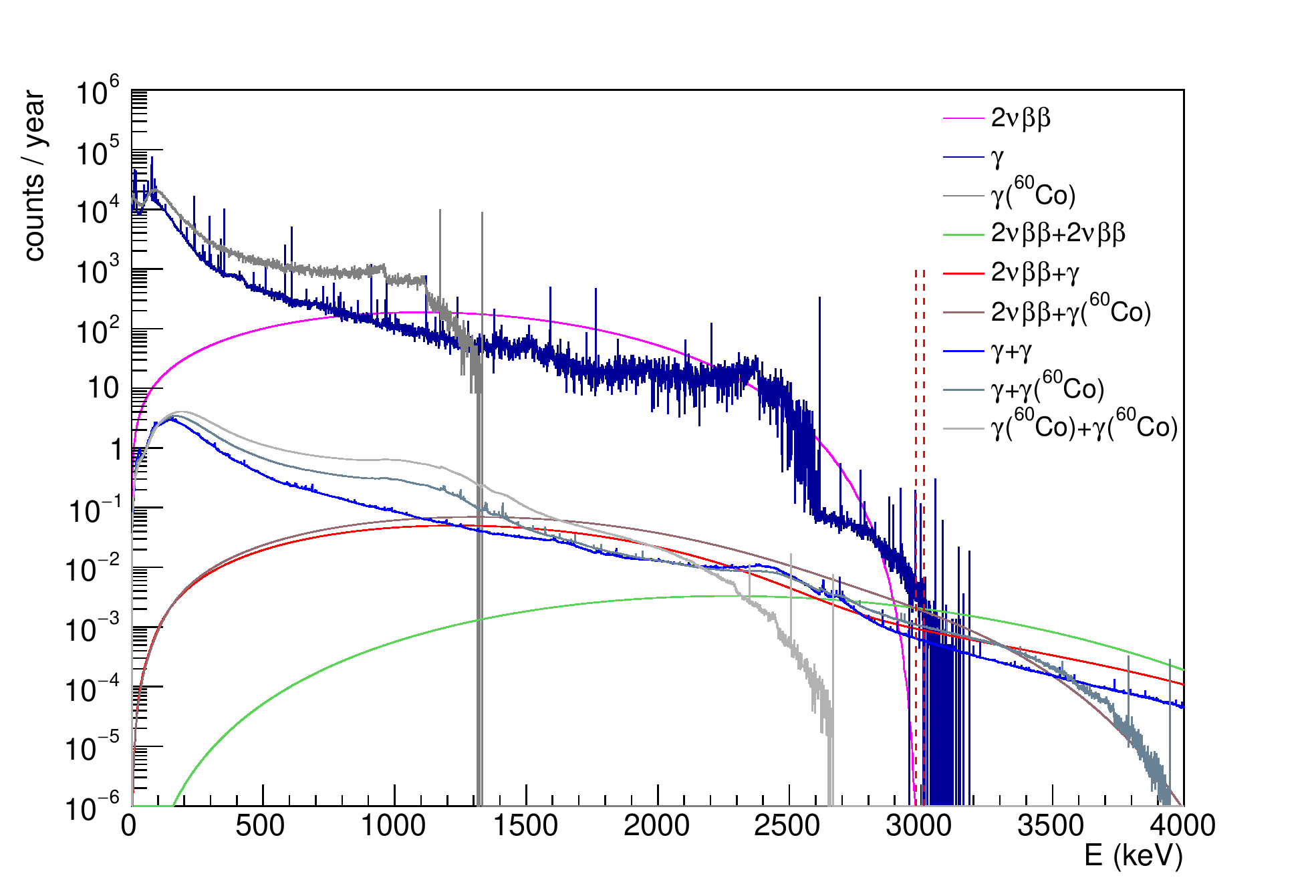}
    \includegraphics[width=0.8\linewidth]{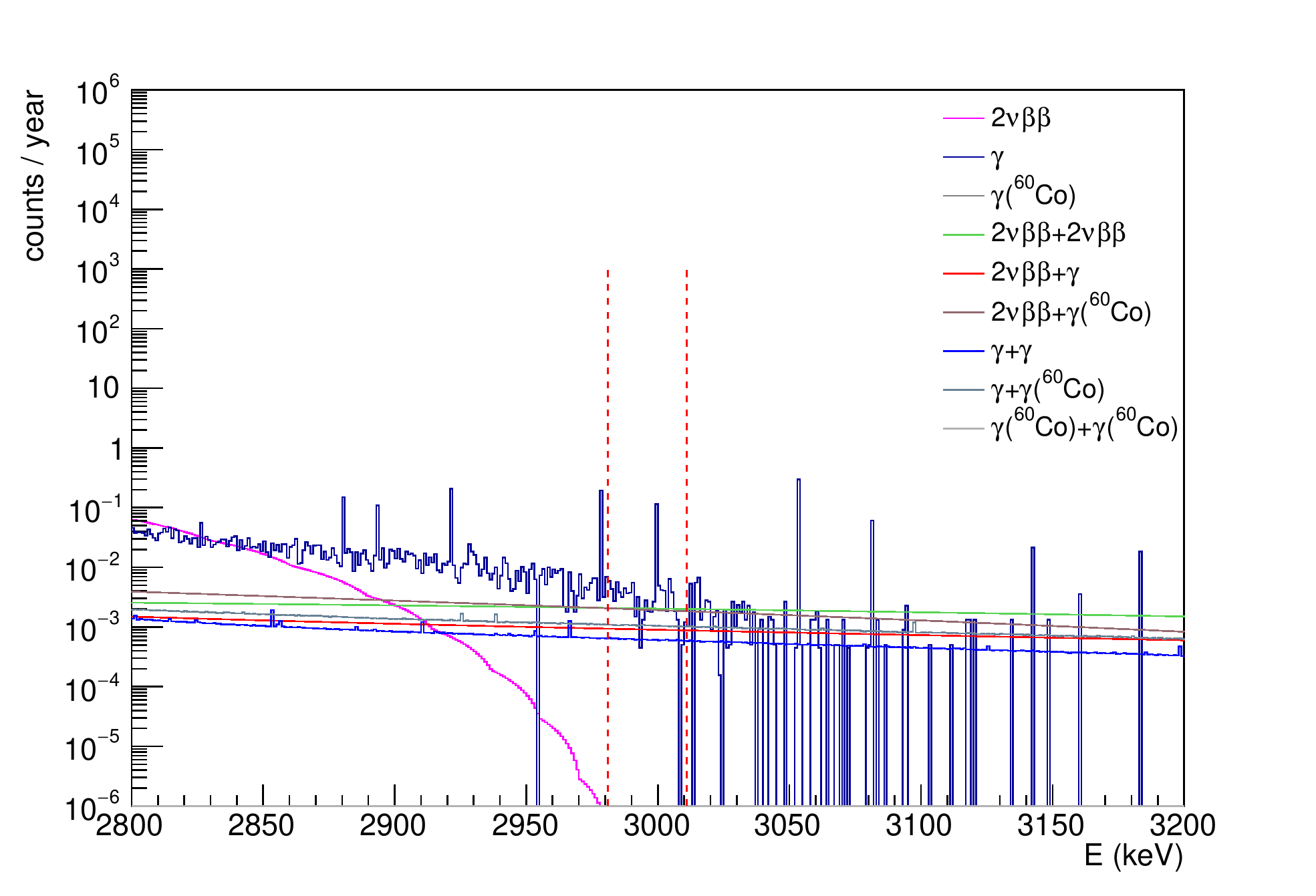}
    \caption{Energy spectra for various single-event and pile-up backgrounds of the \NvDEx-100 experiment with natural \SeF6 gas without further suppression using event topology information. 
}
    \label{fig:pileUpBackgroundNatural}
\end{figure}

\begin{figure}[ht]
    \centering
    \includegraphics[width=0.8\linewidth]{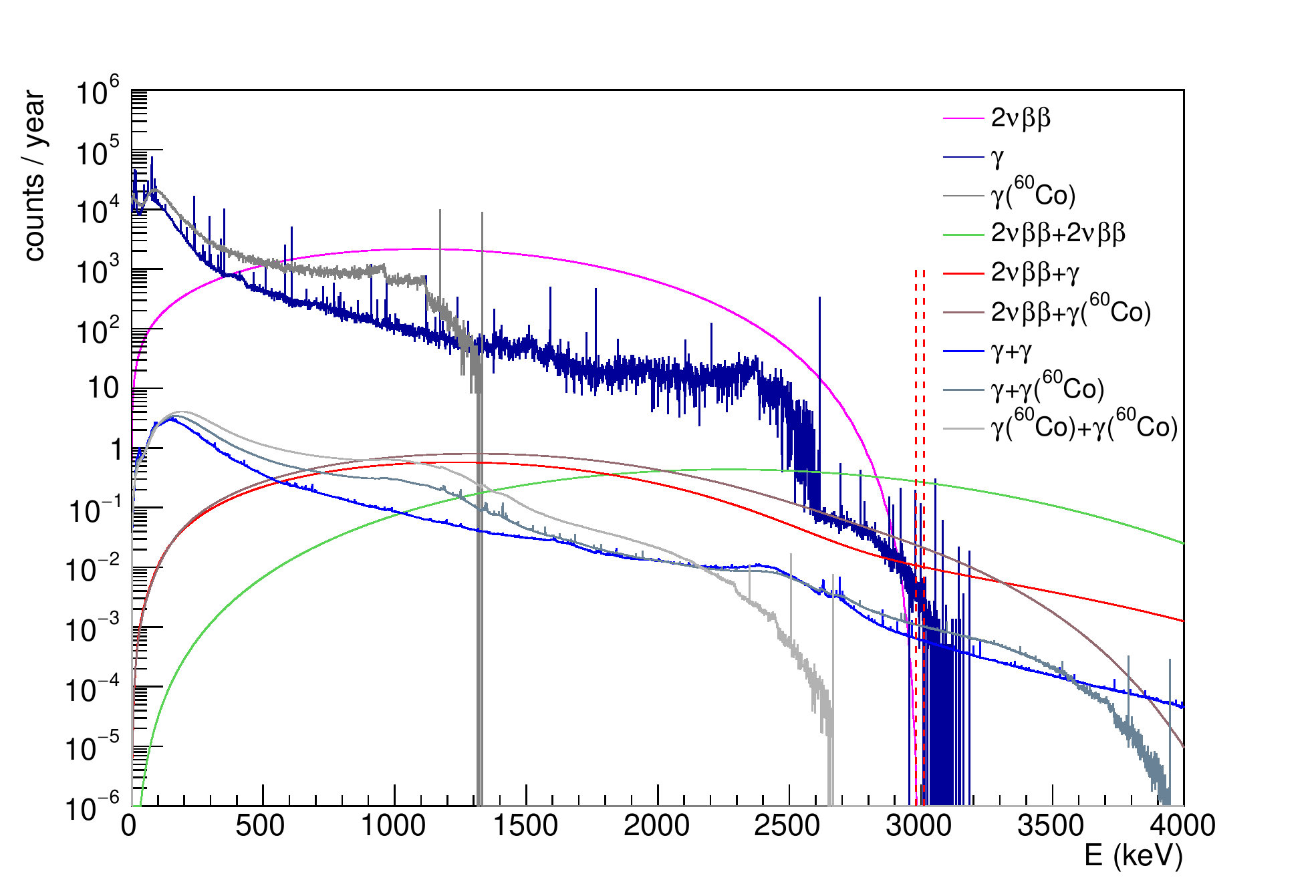}
    \includegraphics[width=0.8\linewidth]{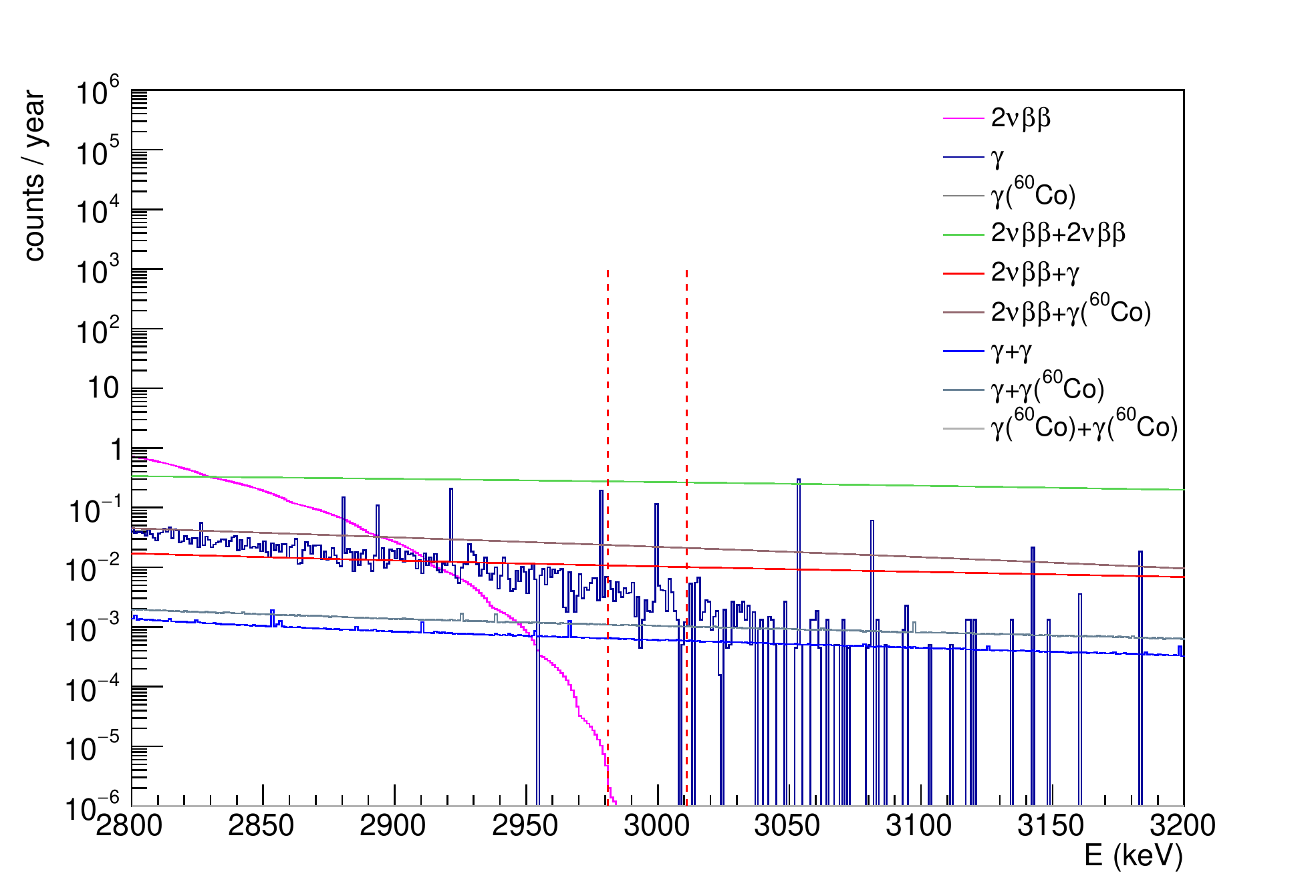}
    \caption{Energy spectra for various single-event and pile-up backgrounds of the \NvDEx-100 experiment with enriched \82SeF6 gas without further suppression using event topology information. 
}
    \label{fig:pileUpBackgroundEnriched}
\end{figure}

Here, we roughly estimate the pile-up event background rate. We assume two events can be separated if they are
10 cm$\times$10 cm$\times$10 cm away when the second event happens. Note that,
10 cm is roughly the size of a \0vbb event in \SeF6 gas at a pressure of 1 MPa.
This is a conservative estimation since events with lower energy dominate the background energy spectrum and are also smaller in size.
With this assumption, we take the single event energy spectra for various backgrounds, do a convolution and a proper normalization, and obtain the pile-up event background energy spectra in Fig. \ref{fig:pileUpBackgroundNatural} and \ref{fig:pileUpBackgroundEnriched} for \NvDEx-100 using natural \SeF6 gas and enriched \82SeF6 gas, respectively.

From the plots, the highest pile-up background component is \2vbb + \2vbb background for the \NvDEx-100 experiment with natural \SeF6 gas, contributing to 0.06 evts/yr in ROI. However, this rate is still lower than that of a single-event natural radiation $\gamma$ background. 

For the \NvDEx-100 experiment with enriched \82SeF6 gas, the \2vbb + \2vbb pile-up background is at the level of eight evts/yr in ROI, which is higher than that of a single-event natural radiation $\gamma$ background.
Thus, pile-up background suppression should be considered in the future for the \NvDEx-100 experiment with enriched \82SeF6 gas. With a full simulation of charge drift, diffusion, and read-out responses, a more careful study of pile-up backgrounds can be conducted in the future, which will be more precise than the current conservative estimation.
Event topology information can also be used with neural networks to reduce the contribution of pile-up events by distinguishing pile-up events from double beta decay events.
For example, a \2vbb + \2vbb pile-up event with two pairs of $\beta$ tracks should look different from a single $\beta\beta$ event, with only one pair of $\beta$ tracks starting from the same position.
If the pile-up backgrounds cannot be suppressed to be smaller than the single-event $\gamma$ background with software alone, adding scintillation light detection at the HV plane side using silicon PM and light guides is also an option to explore in the future. The \SeF6 gas scintillation characteristics must be studied to do this. If needed, other scintillator gases may be mixed into the \82SeF6 gas to increase the scintillation light yield. The HV plane may also need to be changed to a mesh to allow scintillation light to go through. 
If more scintillation light events than drift ion events are observed within a certain drift time, some drift ion events should be pile-up events and should be rejected for further analysis.
With scintillation light read-out, which can easily separate signals from two events happening several ns apart, pile-up background events can be rejected almost completely.

\subsection{Sensitivity Estimation}
\label{sec:sensitivityEstimation}

If \NvDEx-100 is successfully developed and reaches the background level listed above, the dominant background contribution is due to natural radiation $\gamma$, at the level of 0.4 evts / yr in ROI before suppression using event topology information. 

According to studies in ~\cite{Nygren:2018ewr}, the neural network can further suppress the background by a factor of about 10, while keeping a 90\% signal efficiency when event topology information is used. So, the final background level is 0.05 evts / yr in ROI. After five years of operation, the background is about 0.25 events in ROI. Thus, \NvDEx-100 is almost a zero-background experiment. 

\begin{figure}[ht]
    \centering
    \includegraphics[width=8 cm]{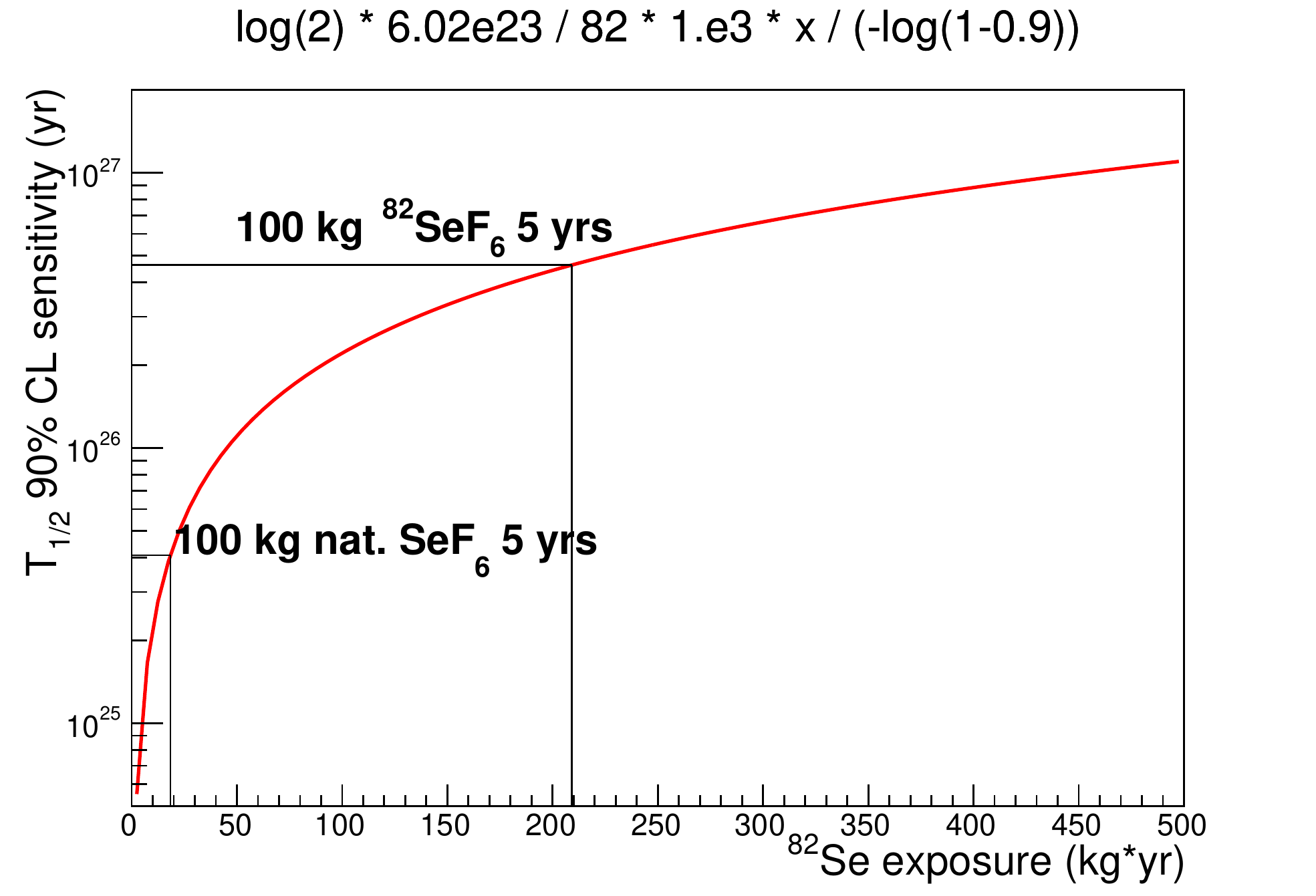}
    \caption{Estimated \NvDEx-100 experiment sensitivity as a function of exposure. 
}
    \label{fig:sensitivity}
\end{figure}

For simplicity, the \NvDEx-100 experiment sensitivity is calculated assuming zero background, as shown in Fig.\ref{fig:sensitivity}. We can see that the $T_{1/2}$ sensitivity can reach $4\times10^{25}$ yrs at 90\% confidence level after five years of operation with 100kg of natural \SeF6 gas. If enriched \82SeF6 gas is used, the $T_{1/2}$ sensitivity can reach $4\times10^{26}$ yrs at 90\% confidence level after five years of running, which is better than the current best $T_{1/2}$ sensitivity of $2.3\times10^{26}$ yrs from the KamLAND-Zen experiment \cite{KamLAND-Zen:2022tow}.

\section{Summary}
\label{sec:summary}

In summary, \NvDEx-100 is a 100-kg scale neutrinoless double beta experiment conducted in the China Jinping Underground Laboratory using a high-pressure \SeF6 \ gas TPC. 
The topmetal-S sensors have been developed to read out drift ion signals in the \NvDEx \ TPC with the electronegative \SeF6 gas. 
All sub-systems of \NvDEx-100, including the pressure chamber and inner copper shielding, TPC field cage, readout plane and data acquisition system, external shielding, gas system, and the negative-pressure clean room, have been completed in the conceptual design and are described in this report.
\NvDEx-100 is being developed, with installation completion at CJPL planned for 2025.
Combining the advantages of the high $Q_{\beta\beta}$ (2.996 MeV) of \Se82 \ and TPC's ability to distinguish signal and background events using their different topological characteristics, \NvDEx-100 \ can achieve a very low background level of 0.05 evts / yr in ROI and high $T_{1/2}$ sensitivity of $4\times10^{25}$ ($4\times10^{26}$) yrs at 90\% confidence level after five  years of operation, using 100 kg of natural \SeF6 (enriched \82SeF6) gas.

\end{document}